\documentstyle[eqsecnum,aps,prb,psfig]{revtex}

\draft
\begin{document}

\title{Microscopic Derivation of the Ginzburg-Landau Equations for a 
d-wave Superconductor}
\author{D.L. Feder and C. Kallin}
\address{Department of Physics and Astronomy, McMaster University, Hamilton,
Ontario L8S 4M1, Canada}
\date{Received \today}
\maketitle
\begin{abstract}

The Ginzburg-Landau (GL) equations for a $d_{x^2-y^2}$ superconductor are
derived
within the context of two microscopic lattice models used to describe the
cuprates: the extended Hubbard model and the Antiferromagnetic-van
Hove model. Both models have pairing on nearest-neighbour links, consistent
with theories for d-wave superconductivity mediated by spin fluctuations.
Analytical results obtained for the extended Hubbard model at low
electron densities and weak-coupling are compared to results reported
previously for a d-wave superconductor in the continuum. The
variation of the coefficients in the GL equations with carrier density,
temperature, and coupling constants are calculated numerically for both
models. The relative importance of anisotropic higher-order terms in the GL 
free energy is investigated, and the implications for experimental observations 
of the vortex lattice are considered.

\end{abstract}
\pacs{74.20.-z, 74.20.De, 74.72.-h}

\widetext

\section{Introduction}

There is mounting experimental evidence to suggest that the high-temperature
cuprate superconductors have an order parameter with unconventional 
symmetry.\cite{revs,scalap} Indeed, recent Josephson interference 
measurements\cite{wollman} are strongly indicative of an order parameter 
with $d_{x^2-y^2}$ (d-wave) symmetry,\cite{sigrist1} which has line nodes along 
$|k_x|=|k_y|$. The linear density of states associated with the resulting 
low-energy excitations
is thought\cite{annett} to account for the linear temperature dependence of the
specific heat\cite{moler} as well as the linear temperature\cite{hardy}
and magnetic field\cite{maeda} dependence of the penetration depth found
for YBa$_2$Cu$_3$O$_{7-\delta}$ (123) and Bi$_2$Sr$_2$CaCu$_2$O$_y$ (2212). 

A number of
experimental results\cite{ma} are consistent only with an order parameter of
combined s-wave and d-wave symmetry. Sigrist and Rice\cite{sigrist2} have shown
that in weakly orthorhombic cuprates (such as 2212 or 123) a small s-wave 
component would be present in addition to a critical d-wave
order parameter. In tetragonal systems (such as the thallium compounds) that 
favour d-wave superconductivity, however, an s-wave component can only be 
nucleated locally near inhomogeneities\cite{volovik} such as domain walls,
impurities,\cite{franz1} or vortices.\cite{pekka,berlinsk,ichioka1,franz2} 
Indeed, Soininen {\em et al.}\cite{pekka} investigated
the structure of an isolated vortex for a d-wave superconductor within 
Bogoliubov-De Gennes theory and found that a non-zero s-wave component is
induced in the vortex core. Their results, interpreted within the context
of the relevant phenomenological Ginzburg-Landau (GL) free energy,\cite{joynt}
imply a non-trivial topological structure for the additional s-wave 
component.\cite{berlinsk} As a consequence, the supercurrent and magnetic field 
distributions for an isolated vortex near $H_{c1}$ exhibit a fourfold
anisotropy in proportion to the magnitude of the s-wave component.\cite{franz2}
In addition, the vortex-lattice structure near $H_{c2}$ 
deviates significantly from the usual triangular Abrikosov lattice, becoming
increasingly oblique with increasing s-wave admixture.\cite{berlinsk,franz2}

It remains uncertain, however, whether the anisotropy in the structures of an
isolated vortex and the vortex lattice is indeed predominantly due to the
admixture of an s-wave component. Ichioka {\em et al.}\cite{ichioka2}
have shown that the inclusion of higher-order d-wave gradients in the GL free
energy can give rise to a fourfold-symmetric current distribution around a
vortex even in the absence of an induced s-wave component. One of the
objectives of the present work is the clarification of this issue. Indeed, we
find that the contributions to anisotropy of a fourth-order gradient term
and the s-wave component are comparable, and tend to compete.

While phenomenological GL theory has been highly successful in
predicting many interesting properties of d-wave superconductors in 
external fields, the relative magnitudes of the various coefficients appearing
in the free energy and their dependence on temperature, filling, and field are
presently unknown. An earlier derivation\cite{ren} of the free energy 
from a continuum
model could not include lattice effects that are believed to be important
in theories of d-wave superconductivity. Consequently, as will be discussed 
later, certain technical
difficulties arose which would not have appeared in the continuum limit of an
appropriate lattice model. In any case, it would be useful to derive the 
GL free energy using models
relevant to the high-$T_c$ oxides. In the present work, the GL
equations are derived microscopically within the context of two such models:
the extended Hubbard and `Antiferromagnetic-van Hove' models. 

The extended Hubbard (EH) model, which includes a nearest-neighbour attraction
in addition to the usual on-site repulsion, is one of the simplest lattice 
models which allows for a d-wave superconducting instability. Pairing occurs
along nearest-neighbour links, appropriate for theories where d-wave
superconductivity is mediated by antiferromagnetic fluctuations (see 
Ref.\ \onlinecite{scalap} and references therein). It has been 
employed in several analytical\cite{micnas} and 
numerical\cite{pekka,micnas,wang}
investigations of d-wave superconductivity. The EH model has
recently been shown,\cite{naz} however, to favour d-wave superconductivity 
only in a 
very small parameter space, preferring a phase separated or spin-density wave
state.

The `Antiferromagnetic-van Hove' (AvH) model\cite{dagotto} strongly favours 
d-wave superconductivity while incorporating the coexisting 
antiferromagnetic correlations observed in NMR,\cite{imai} neutron 
scattering,\cite{aeppli}
and angle-resolved photoemission spectroscopy (ARPES)\cite{aebi} experiments.
High transition temperatures are obtained in the model due to the presence 
of a van Hove
singularity in the hole density of states near the Fermi energy. An extended,
flat band near $({\pi\over 2},{\pi\over 2})$ in momentum space is consistent 
with numerical investigations of a single 
hole propagating through an antiferromagnetic background\cite{bos}, and
with experimental evidence.\cite{dessau} In
the AvH model, holes are constrained to move within a single sublattice of
a uniform antiferromagnetic background in order to minimize frustration. The
hopping parameters are chosen to best fit the quasiparticle dispersion for YBCO
measured using ARPES.\cite{abrikosov}

In Section II, the Ginzburg-Landau equations for the gap functions and 
supercurrent are derived microscopically for both the EH and AvH lattice 
models using a finite-temperature Green function method. The
relations defining the transition temperatures are investigated in Section
III. It is found that only a d-wave transition is favoured for the AvH model;
the d-wave transition temperature $T_d\sim 100 $ K is consistent with the 
high-temperature (high-$T_c$) oxides.
The EH model, in contrast, can have either an s-wave or d-wave instability.
S-wave is favoured at low electron densities while d-wave is favoured either
at high densities or at lower densities with strong on-site repulsion.
The equations
for $T_s$ (the s-wave transition temperature) and $T_d$ are found analytically 
in the limit of weak-coupling and
low electron densities. The corresponding analytical solutions for the AvH 
model are difficult
to obtain due to the complicated angular dependence of the AvH dispersion.
The GL free energy is derived for both models in Section IV. The coefficients 
of the GL equations are found
analytically for the EH model in the same limit described above. The 
coefficients are calculated numerically for the EH model
and for the AvH model near
optimal doping. In section V, we summarize our results and discuss the 
experimental implications of the GL equations we have derived.

\section{The lattice GL equations}

The Hamiltonians for the extended Hubbard (EH) and antiferromagnetic-van Hove
(AvH) models are respectively:

\begin{equation}
H^{\rm EH}=-t\sum_{\langle ij\rangle\sigma}c^{\dag}_{i\sigma}
c^{\vphantom{\dag}}_{j\sigma}
e^{i{2\pi\over\phi_{\circ}}\int_j^i{\bf A}\cdot d{\bf l}}
-\mu\sum_{i\sigma}n_{i\sigma}
+V_0\sum_in_{i\uparrow}n_{i\downarrow}-{1\over 2}
\sum_{\langle ij\rangle\sigma\sigma^{\prime}}
V_{ij}n_{i\sigma}n_{j\sigma^{\prime}};
\end{equation}

\begin{equation}
H^{\rm AvH}=t_{11}\sum_{\langle\langle ij\rangle\rangle\sigma}
c^{\dag}_{i\sigma}c^{\vphantom{\dag}}_{j\sigma}
e^{i{2\pi\over\phi_{\circ}}\int_j^i{\bf A}\cdot d{\bf l}}
+t_{20}\sum_{\langle\langle\langle ij\rangle\rangle\rangle\sigma}
c^{\dag}_{i\sigma}c^{\vphantom{\dag}}_{j\sigma}
e^{i{2\pi\over\phi_{\circ}}\int_j^i{\bf A}\cdot d{\bf l}}
-\mu\sum_{i\sigma}n_{i\sigma}
-{V\over 2}\sum_{\langle ij\rangle\sigma\sigma'}
n_{i\sigma}n_{j\sigma'},
\end{equation}

\noindent where $n_{i\sigma}=c^{\dag}_{i\sigma}c^{\vphantom{dag}}_{i\sigma}$,
{\bf A} is the vector potential associated with the external magnetic field, 
$\phi_{\circ}=hc/e$ is the flux quantum, and $\mu$ is the chemical potential 
included to fix the density. In the EH model the carriers are electrons, and
positive $V_0$ and $V_{ij}$ imply on-site 
repulsion and nearest-neighbour attraction, respectively. The superconducting
carriers in the AvH model
are holes propagating through the antiferromagnetic background
of the undoped `parent' state. Second and 
third nearest-neighbour hopping parameters are respectively
$t_{11}=0.04125$ eV and $t_{20}=0.02175$ eV.\cite{dagotto} The absence of 
nearest-neighbour hopping in the AvH model reflects the restricted Hilbert 
space of the carriers; holes are constrained to move within a single spin
sublattice in order to minimize frustration and preserve antiferromagnetic 
correlations. The values
of $t_{11}$ and $t_{20}$ are chosen to result in a large density of 
states near the bottom of the hole band, located at $(\pi/2,\pi/2)$ in momentum 
space. The coefficient of the
nearest-neighbour attraction $V=0.075$ eV is chosen to yield a d-wave 
transition temperature $T_d\sim 100$ K at optimal doping ($\mu\approx -0.075$ 
eV, or hole density $\langle n\rangle\sim 0.2$), consistent with the high-$T_c$ 
oxides.\cite{dagotto} 

If the lattice sites $i$ and $j$ are nearest-neighbours, we can write the 
mean-field EH Hamiltonian 

\begin{eqnarray}
H^{\rm EH}_{\rm eff}(B)&=&-t\sum_{{\bf r},\vec{\delta},\sigma}c_{\sigma}^{\dag}
({\bf r}+\vec{\delta})c^{\vphantom{\dag}}_{\sigma}({\bf r})e^{i\phi_{\delta}}
-\mu\sum_{{\bf r},\sigma}c^{\dag}_{\sigma}({\bf r})
c^{\vphantom{\dag}}_{\sigma}({\bf r})
+\sum_{\bf r}\left[\Delta^*_{\circ}({\bf r})
c_{\downarrow}({\bf r})c_{\uparrow}({\bf r}) + H.c.\right]\nonumber \\
&-&{1\over 2}\sum_{{\bf r},\vec{\delta}}\big[\Delta^*_{\delta}({\bf r})c_{\downarrow}
({\bf r})c_{\uparrow}({\bf r}+\vec{\delta})
-\Delta^*_{\delta}({\bf r})c_{\uparrow}({\bf r})
c_{\downarrow}({\bf r}+\vec{\delta}) + H.c.\big],
\end{eqnarray}

\noindent where 

\begin{equation}
\phi_{\delta}={2\pi\over\phi_{\circ}}\int_{\bf r}^{{\bf r}+\vec{\delta}}{\bf A}
\cdot d\,{\bf l},
\end{equation}

\noindent and $\vec{\delta}=\pm\hat{x},\pm\hat{y}$ (the lattice constant is 
taken to 
be unity for convenience). The `on-site' and nearest-neighbour gap functions
are defined as follows:

\begin{equation}
\Delta_{\circ}({\bf r})\equiv V_0\langle c_{\downarrow}({\bf r})
c_{\uparrow}({\bf r})\rangle;
\end{equation}

\begin{eqnarray}
\Delta_{\delta}({\bf r})&\equiv& V_{\delta}\langle c_{\downarrow}({\bf r})
c_{\uparrow}({\bf r}+\vec{\delta})\rangle\nonumber \\
&\equiv& -V_{\delta}\langle 
c_{\uparrow}({\bf r})c_{\downarrow}({\bf r}+\vec{\delta})\rangle ,
\end{eqnarray}

\noindent assuming the existence of pairing in only the spin-singlet channel,
in accordance with experimental results for the cuprate 
superconductors.\cite{singlet} The mean-field Hamiltonian for the AvH model is 
written:

\begin{eqnarray}
H^{\rm AvH}_{\rm eff}(B)&=&t_{11}\sum_{{\bf r},\vec{\delta}_{11},\sigma}
c_{\sigma}^{\dag}({\bf r}+\vec{\delta}_{11})c^{\vphantom{\dag}}_{\sigma}
({\bf r}) e^{i\phi_{\vec{\delta}_{11}}}
+t_{20}\sum_{{\bf r},\vec{\delta}_{20},\sigma}
c_{\sigma}^{\dag}({\bf r}+\vec{\delta}_{20})c^{\vphantom{\dag}}_{\sigma}
({\bf r}) e^{i\phi_{\vec{\delta}_{20}}}
-\mu\sum_{{\bf r},\sigma}c^{\dag}_{\sigma}({\bf r})
c^{\vphantom{\dag}}_{\sigma}({\bf r})\nonumber \\
&-&{1\over 2}\sum_{{\bf r},\vec{\delta}}\big[\Delta^*_{\delta}({\bf r})
c_{\downarrow} ({\bf r})c_{\uparrow}({\bf r}+\vec{\delta})+H.c.\big],
\end{eqnarray}

\noindent where ${\bf r}=m\hat{r}_1+n\hat{r}_2$, such that 
$\hat{r}_1\equiv\hat{x}+\hat{y}$ and
$\hat{r}_2\equiv\hat{x}-\hat{y}$ are primitive vectors of a single sublattice,
and each lattice site has the two-point basis $\hat{0}$, $\hat{x}$. Then,
$\vec{\delta}_{11}=\pm\hat{r}_1,\pm\hat{r}_2$,
$\vec{\delta}_{20}=\pm(\hat{r}_1+\hat{r}_2),\pm(\hat{r}_1-\hat{r}_2)$, and
$\vec{\delta}=\pm\hat{x},\pm\hat{y}$. 
Throughout the remainder of this section calculations will be presented 
within the
context of the EH model. Comparison of the above Hamiltonians indicates that 
analogous results for the AvH model can be obtained at any stage by 
eliminating the on-site gap function, reversing the sign of the kinetic term 
(keeping in mind the holes hop along second and third-neighbour links), and 
setting $V_{\delta}=V/2$. 

The Gor'kov equations can then be derived in the standard manner:\cite{gorkov}

\begin{equation}
{\cal G}({\bf r},{\bf r}^{\prime},\omega_n)=\tilde{\cal G}^{\circ}
({\bf r},{\bf r}^{\prime},\omega_n)+\sum_{{\bf r}''}\tilde{\cal G}^{\circ}
({\bf r},{\bf r}'',\omega_n)\big[\Delta_{\circ}({\bf r}'')
-\sum_{\vec{\delta}}\Delta_{\delta}({\bf r}'')
\hat{P}_{\delta}({\bf r}'')\big]{\cal F}^{\dag}({\bf r}'',{\bf r}',\omega_n);
\label{gorkov3}
\end{equation}

\begin{equation}
{\cal F}^{\dag}({\bf r},{\bf r}^{\prime},\omega_n)=
-\sum_{{\bf r}''}\tilde{\cal G}^{\circ}
({\bf r}'',{\bf r},-\omega_n)\big[\Delta^*_{\circ}({\bf r}'')
-\sum_{\vec{\delta}}\Delta^*_{\delta}({\bf r}'')
\hat{P}_{\delta}({\bf r}'')\big]{\cal G}({\bf r}'',{\bf r}',\omega_n),
\label{gorkov4}
\end{equation}

\noindent where $\hat{P}_{\delta}({\bf r})X({\bf r})\equiv X({\bf r}
+\vec{\delta})$ is the kinetic energy operator,
${\cal G}$ and ${\cal F}^{\dag}$ are finite-temperature single-particle
and anomalous Green functions respectively, and $\tilde{\cal G}^{\circ}$ is 
the normal-state Green function in the presence of the external field. 
The Matsubara frequencies are $\omega_n\equiv\pi T(2n+1)$. The anomalous Green
function ${\cal F}^{\dag}$ is related to the gap functions:

\begin{eqnarray} 
\Delta^*_{\circ}({\bf r})&=&TV_0\sum_{\omega_n}{\cal F}^{\dag}(
{\bf r},{\bf r},\omega_n),\label{d0}\\
\Delta^*_{\delta}({\bf r})&=&TV_{\delta}\sum_{\omega_n}{\cal F}^{\dag}({\bf r},
{\bf r+\vec{\delta}},\omega_n),\label{dsd}
\end{eqnarray}

\noindent Arbitrarily close to the superconducting critical temperature $T_c$,
the ratios $\Delta_{\alpha}/T_c\ll 1$ or identically 
zero ($\vec{\alpha}=0,\vec{\delta}$). 
Iterating Eqs.\ (\ref{gorkov3}) and (\ref{gorkov4}) up to third order in the 
gap functions, and making use of the conditions (\ref{d0}) and (\ref{dsd}), the 
self-consistent equations for the gap functions are immediately obtained:

\begin{eqnarray}
{1\over V_{\alpha}}\Delta^*_{\alpha}({\bf r})&=&-\sum_{{\bf r}'',\omega_n}
\tilde{\cal G}^{\circ}({\bf r}'',{\bf r},-\omega_n)\Delta^*({\bf r}'')
\tilde{\cal G}^{\circ}({\bf r}'',{\bf r}+\vec{\alpha},\omega_n)\nonumber \\
&+&\sum_{{\bf r}'',{\bf r}_1,{\bf r}_2,\omega_n}
\tilde{\cal G}^{\circ}({\bf r}'',{\bf r},-\omega_n)\Delta^*({\bf r}'')
\tilde{\cal G}^{\circ}({\bf r}'',{\bf r}_1,\omega_n)\Delta({\bf r}_1)
\tilde{\cal G}^{\circ}({\bf r}_2,{\bf r}_1,-\omega_n)\Delta^*({\bf r}_2)
\tilde{\cal G}^{\circ}({\bf r}_2,{\bf r}+\vec{\alpha},\omega_n),
\label{gap_eq}
\end{eqnarray}

\noindent where 

\begin{equation}
\Delta^*({\bf x})\equiv\Delta^*_{\circ}({\bf x})-\sum_{\vec{\delta}}
\Delta^*_{\delta}({\bf x})\hat{P}_{\delta}({\bf x}),
\end{equation}

\noindent and $\vec{\alpha}=\hat{0}$, $\pm\hat{x}$, or $\pm\hat{y}$.

In the strong type-II limit, appropriate for the high-$T_c$ oxides, the 
penetration depth $\lambda(T)$ exceeds
the coherence length $\xi(T)$ and all other length scales. The
single-particle Green function is then approximately translationally 
invariant:

\begin{eqnarray}
\tilde{\cal G}^{\circ}({\bf r},{\bf r}',\omega_n)&\approx&{\cal G}^{\circ}
({\bf r}-{\bf r}',\omega_n)e^{-i{2\pi\over\phi_{\circ}}\int_{\bf r}^{{\bf r}'}
{\bf A}\cdot d{\bf l}}\nonumber \\
&\approx&{\cal G}^{\circ}({\bf r}-{\bf r}',\omega_n)e^{i{2\pi\over\phi_{\circ}}
{\bf A}({\bf r})\cdot ({\bf r}-{\bf r}')},
\end{eqnarray}

\noindent where ${\cal G}^{\circ}({\bf r}-{\bf r}',\omega_n)$ is the 
normal-state lattice Green function in the absence of an external field:

\begin{equation}
{\cal G}^{\circ}({\bf x},\omega_n)=\sum_{\bf k}{e^{i{\bf k}\cdot{\bf x}}\over
i\omega_n-\xi_k},
\label{green}
\end{equation}

\noindent and the sum is over wavevectors in the first Brillouin zone. The 
dispersion relations for the EH and AvH models are respectively

\begin{eqnarray}
\xi^{\rm EH}_k&=&-2t(\hbox{cos}k_x+\hbox{cos}k_y)-\mu\label{disp_eh} \\
\xi^{\rm AvH}_k&=&2t_{11}(\hbox{cos}k_1+\hbox{cos}k_2)+4t_{20}\hbox{cos}k_1
\hbox{cos}k_2-\mu,\label{disp_avh}
\end{eqnarray}

\noindent where $k_1$ and $k_2$ are reciprocal vectors of a given sublattice. 
Assuming the gap functions vary slowly compared with the characteristic
length scale $k_F$ of the single-particle Green function (\ref{green}),
Eq.\ (\ref{gap_eq}) can be expanded up to fourth-order in lattice derivatives.
The justification for keeping higher-order gradient terms will be addressed in 
section V.  The GL equations for the gap functions in the EH model can then be 
written:

\begin{eqnarray}
\Delta^*_{\alpha}({\bf r})&=&-TV_{\alpha}\sum_{\omega_n}\sum_{\vec{\alpha}'}
(-1)^{\alpha'}
\sum_{m,n}{\cal G}^{\circ}\big(m\hat{x}+n\hat{y},-\omega_n\big)
{\cal G}^{\circ}\big(m\hat{x}+n\hat{y}+\vec{\alpha}'-\vec{\alpha},\omega_n
\big) e^{-i{2\pi\over\phi_0}{\bf A}({\bf r})\cdot (\vec{\alpha}-\vec{\alpha}')}
\nonumber \\
& &\Bigg\{ 1-\epsilon_{x,1}^{\rm EH}\left(\hat{x}\Pi_{x}\right)^2
-\epsilon_{y,1}^{\rm EH}\left(\hat{y}\Pi_{y}\right)^2
-\epsilon_{x,2}^{\rm EH}\left(\hat{x}\Pi_x\right)^4
-\epsilon_{y,2}^{\rm EH}\left(\hat{y}\Pi_y\right)^4
-\epsilon_{xy,2}^{\rm EH}\left(\hat{x}\Pi_x\right)^2
\left(\hat{y}\Pi_y\right)^2\Bigg\}\Delta^*_{\alpha'}({\bf r})\nonumber \\
&+&TV_{\alpha}\sum_{{\bf k},\omega_n}{e^{i{\bf k}\cdot\vec{\alpha}}
e^{-i{2\pi\over\phi_0}{\bf A}({\bf r})\cdot\vec{\alpha}}
\over(\omega_n^2+\xi_k^2)^2}\bigg\{|\Delta_{\circ}({\bf r})|^2
\Delta^*_{\circ}({\bf r})-2|\Delta_{\circ}({\bf r})|^2M({\bf r})
+2|M({\bf r})|^2\Delta^*_{\circ}({\bf r})\nonumber \\
& &\qquad\quad -{\Delta^*}^2_{\circ}({\bf r})M^*({\bf r})
+\Delta_{\circ}({\bf r})M^2({\bf r})-|M({\bf r})|^2M({\bf r})\bigg\},
\label{gaps_eh}
\end{eqnarray}

\noindent where 

\begin{eqnarray}
i\left(\hat{z}\Pi_z\right)\Delta_{-z}({\bf r})&\equiv&
e^{i{4\pi\over\phi_0}A_z({\bf r})}\Delta_{z}({\bf r})-\Delta_{-z}({\bf r})
\label{def_grad}\\
&\approx&\left[\left(\hat{z}\hat{\Delta}_z\right)+i{4\pi\over\phi_0}
A_z({\bf r})\right]\Delta_{-z}({\bf r}),
\end{eqnarray}

\noindent with the forward difference operator defined by

\begin{eqnarray}
\left(\hat{z}\hat{\Delta}_{z}\right)\Delta_{-z}({\bf r})&\equiv&
\Delta_{-z}({\bf r}+\hat{z})-\Delta_{-z}({\bf r})\nonumber \\
&=&\Delta_z({\bf r})-\Delta_{-z}({\bf r}),
\end{eqnarray}

\noindent and

\begin{equation}
M({\bf r})\equiv\sum_{\vec{\delta}'=\pm\hat{x},\pm\hat{y}}\Delta^*_{\delta'}
({\bf r})e^{i{\bf k}\cdot\vec{\delta}'}e^{i{2\pi\over\phi_0}{\bf A}({\bf r})
\cdot\vec{\delta}'}.
\end{equation}

\noindent The coefficients of the gradient terms have been defined in the 
Appendix for clarity. Note that all gap functions are taken at the same 
point ${\bf r}$ in the last term of Eq.\ (\ref{gaps_eh}); this condition
will be relaxed in Section V. The gradient terms
with odd powers of $\hat{x}\Pi_x$ and/or $\hat{y}\Pi_y$ vanish for tetragonal
systems. The GL equations for the gap functions in the AvH model are given by:

\begin{eqnarray}
\Delta^*_{\delta}({\bf r})&=&{TV\over 2}\sum_{\omega_n}\sum_{\vec{\delta}'}
\sum_{m,n}{\cal G}^{\circ}\big(m\hat{r}_1+n\hat{r}_2,-\omega_n\big)
{\cal G}^{\circ}\big(m\hat{r}_1+n\hat{r}_2+\vec{\delta}'-\vec{\delta},\omega_n
\big)e^{-i{2\pi\over\phi_0}{\bf A}({\bf r})\cdot (\vec{\delta}-\vec{\delta}')}
\nonumber \\
& &\Bigg\{
1-\epsilon_{x,1}^{\rm AvH}\left(\hat{x}\Pi_{x}
\right)^2-\epsilon_{y,1}^{\rm AvH}\left(\hat{y}\Pi_{y}\right)^2
-\epsilon_{x,2}^{\rm AvH}\left(\hat{x}\Pi_x\right)^4
-\epsilon_{y,2}^{\rm AvH}\left(\hat{y}\Pi_y\right)^4
-\epsilon_{xy,2}^{\rm AvH}\left(\hat{x}\Pi_x\right)^2\left(\hat{y}\Pi_y\right)^2
\Bigg\}\Delta^*_{\delta'}({\bf r})\nonumber \\
&-&{TV\over 2}\sum_{{\bf k},\omega_n}{e^{i{\bf k}\cdot\vec{\delta}}
e^{-i{2\pi\over\phi_0}{\bf A}({\bf r})\cdot\vec{\delta}}
\over(\omega_n^2+\xi_k^2)^2}|M({\bf r})|^2M({\bf r}),
\label{gaps_avh}
\end{eqnarray}

\noindent where we have used

\begin{eqnarray}
\left(\hat{r}_1\Pi_{r_1}\right)&\equiv&\left(\hat{x}\Pi_x\right)
\left(\hat{y}\Pi_y\right)+\left(\hat{x}\Pi_x\right)+\left(\hat{y}
\Pi_y\right);\\
\left(\hat{r}_2\Pi_{r_2}\right)&\equiv& -\left(\hat{x}\Pi_x\right)
\left(\hat{y}\Pi_y\right)+\left(\hat{x}\Pi_x\right)-\left(\hat{y}
\Pi_y\right),
\end{eqnarray}

\noindent and the coefficients $\epsilon$ are given in the Appendix.

The current operator for the EH model is

\begin{eqnarray}
{\bf j}^{\rm EH}({\bf r})&=&it\sum_{\vec{\delta},\sigma}\vec{\delta}
e^{i\phi_{\delta}}c_{\sigma}^{\dag}
({\bf r}+\vec{\delta})c_{\sigma}^{\vphantom{dag}}({\bf r})\nonumber \\
&=&-2tT\sum_{\vec{\delta},\omega_n}\Pi_{\delta}'({\bf r}'){\cal G}({\bf r},
{\bf r}',\omega_n)\Big|_{{\bf r}={\bf r}'}\quad ,\qquad i\Pi_{\delta}'
({\bf r})f({\bf r})\equiv e^{i{2\pi\over\phi_0}{\bf A}({\bf r})\cdot
\vec{\delta}}f({\bf r}+\vec{\delta})-f({\bf r})\label{cur_lat} \\
&\rightarrow&-2t\left[\vec{\Pi}'^*({\bf r})+\vec{\Pi}'({\bf r}')\right]
T\sum_{\omega_n}{\cal G}({\bf r},{\bf r}',\omega_n)\Big|_{{\bf r}={\bf r}'}
\quad ,\qquad \vec{\Pi}'({\bf r})\equiv -i\vec{\nabla}+{2\pi\over\phi_0}
{\bf A}({\bf r}),\label{cur_cont}
\end{eqnarray}

\noindent where (\ref{cur_cont}) represents the continuum limit of the lattice
current. Iterating (\ref{gorkov3}) and (\ref{gorkov4}) to second order in gap
functions and making the same approximations employed above, (\ref{cur_lat})
becomes:

\begin{eqnarray}
{\bf j}^{\rm EH}({\bf r})=2tT\sum_{{\scriptstyle\vec{\delta},\omega_n\atop
\scriptstyle\vec{\alpha}_1,\vec{\alpha}_2}\atop\scriptstyle
{\bf R}_1,{\bf R}_2}& &e^{i{2\pi\over\phi_0}{\bf A}({\bf r})\cdot(\vec{\alpha}_2
-\vec{\alpha}_1)}(-1)^{|\alpha_1|}(-1)^{|\alpha_2|}{\cal G}^{\circ}(-{\bf R}_1,
\omega_n){\cal G}^{\circ}({\bf R}_2-{\bf R}_1-\vec{\alpha}_1,-\omega_n)
\hat{\Delta}_{-\delta}({\bf R}_2){\cal G}^{\circ}({\bf R}_2+\vec{\alpha}_2,
\omega_n)\nonumber \\
& &\cdot\left\{\Delta^*_{\alpha_2}({\bf r})\left[m\left(\hat{x}\Pi_x
\right)^*+n\left(\hat{y}\Pi_y\right)^*\right]\Delta_{\alpha_1}({\bf r})
+\Delta_{\alpha_1}({\bf r})\left[m'\left(\hat{x}\Pi_x\right)
+n'\left(\hat{y}\Pi_y\right)\right]\Delta^*_{\alpha_2}({\bf r})\right\},
\label{cur_eh}
\end{eqnarray}

\noindent where ${\bf R}_1=m\hat{x}+n\hat{y}$ 
and ${\bf R}_2=m'\hat{x}+n'\hat{y}$. A similar expression for the AvH model
can be obtained by setting $t\rightarrow -t_{11},-t_{20}$ 
with $\vec{\delta}\rightarrow\vec{\delta}_{11},\vec{\delta}_{20}$ and
${\bf R}_1=m\hat{r}_1+n\hat{r}_2$, ${\bf R}_2=m'\hat{r}_1+n'\hat{r}_2$,
and making
the replacements $(\hat{x}\Pi_x)\rightarrow(\hat{x}\Pi_x)+(\hat{y}\Pi_y)$ and
$(\hat{y}\Pi_y)\rightarrow(\hat{x}\Pi_x)-(\hat{y}\Pi_y)$.

The various integrals and sums appearing in the Ginzburg-Landau equations for
the gaps functions (\ref{gaps_eh}), (\ref{gaps_avh}), and the current 
(\ref{cur_eh}) can, in general, be determined only numerically. There is,
however, one case which is analytically tractable: the EH model at low 
electron densities and weak to intermediate coupling. For $V_{\delta}=0$ 
and $V_0\rightarrow -V_0$, this limit would correspond to ordinary BCS theory.

\section{Determination of $T_c$}

At temperatures sufficiently near $T_c$, we can linearize the gap equations 
(\ref{gaps_eh}) and (\ref{gaps_avh}). Making use of the definition of the 
normal-state Green function (\ref{green}), we obtain for the EH model:

\begin{eqnarray}
\Delta^*_{\circ}({\bf r})&=&TV_0\sum_{\omega_n,{\bf k}}{
2a_k\Delta^*_s({\bf r})-\Delta^*_{\circ}({\bf r})\over\omega_n^2+\xi_k^2}
\label{delta0_eh}, \\
\Delta^*_s({\bf r})&=&{TV_1\over 2}\sum_{\omega_n,{\bf k}}{a_k\left[
2a_k\Delta^*_s({\bf r})-\Delta^*_{\circ}({\bf r})\right]
\over\omega_n^2+\xi_k^2},\label{deltas_eh} \\
\Delta^*_d({\bf r})&=&TV_1\sum_{\omega_n,{\bf k}}{b_k^2
\Delta^*_d({\bf r})\over\omega_n^2+\xi_k^2},\label{deltad_eh}
\end{eqnarray}

\noindent where $a_k=\hbox{cos}k_x+\hbox{cos}k_y$ and
$b_k=\hbox{cos}k_x-\hbox{cos}k_y$. The s-wave and d-wave gap functions are 
related to the bond gap functions through the gauge-invariant 
definitions:\cite{invert}

\begin{eqnarray}
\Delta_s({\bf r})&\equiv&{1\over 4}\left[
e^{-i{2\pi\over\phi_0}A_x({\bf r})}\Delta_x({\bf r})
+e^{i{2\pi\over\phi_0}A_x({\bf r})}\Delta_{-x}({\bf r})
+e^{-i{2\pi\over\phi_0}A_y({\bf r})}\Delta_y({\bf r})
+e^{i{2\pi\over\phi_0}A_y({\bf r})}\Delta_{-y}({\bf r})\right],\label{def_s} \\
\Delta_d({\bf r})&\equiv&{1\over 4}\left[
e^{-i{2\pi\over\phi_0}A_x({\bf r})}\Delta_x({\bf r})
+e^{i{2\pi\over\phi_0}A_x({\bf r})}\Delta_{-x}({\bf r})
-e^{-i{2\pi\over\phi_0}A_y({\bf r})}\Delta_y({\bf r})
-e^{i{2\pi\over\phi_0}A_y({\bf r})}\Delta_{-y}({\bf r})\right].\label{def_d} \\
\end{eqnarray}

The equations that determine the s-wave and d-wave 
transition temperatures $T_s$ and $T_d$ are immediately

\begin{equation}
\left[V_1I_2(T_s)-2\right]\left[V_0I_0(T_s)+2\right]=V_0V_1I_1^2
(T_s);
\label{ts_cond}
\end{equation}

\begin{equation}
I_3(T_d)={2\over V_1},
\label{td_cond}
\end{equation}

\noindent where 

\begin{eqnarray}
I_n(T_s)&\equiv&\sum_{\bf k}{a_k^n\over\xi_k}\hbox{tanh}\left({\xi_k
\over 2T_s}\right),\;n=0,1,2;\label{in} \\
I_3(T_d)&\equiv&\sum_{\bf k}{b_k^2\over\xi_k}\hbox{tanh}\left({\xi_k
\over 2T_d}\right).\label{i3}
\end{eqnarray}

\noindent It is clear from Eq. (\ref{ts_cond}) that if $V_1=0$, no s-wave 
superconducting instability can occur for positive temperature.

The corresponding equations for the AvH model are

\begin{eqnarray}
1&=&{V\over 4}\sum_{\bf k}{1+\hbox{cos}k_1\hbox{cos}k_2
+\hbox{cos}k_1+\hbox{cos}k_2\over\xi_k}\hbox{tanh}\left({\xi_k\over 2T_s}
\right),\label{deltas_avh} \\
1&=&{V\over 4}\sum_{\bf k}{1+\hbox{cos}k_1\hbox{cos}k_2
-\hbox{cos}k_1-\hbox{cos}k_2\over\xi_k}\hbox{tanh}\left({\xi_k\over 2T_d}
\right).\label{deltad_avh}
\end{eqnarray}

\noindent While these equations are not analytically tractable, they have been 
shown numerically\cite{dagotto} to strongly favour a d-wave 
transition temperature for all relevant hole concentrations. Optimal doping at 
approximately 25\% filling (in the hole representation, 
where $\langle n_h\rangle$ is
calculated at $T_c$) yields a value of $T_d\sim 100$ K.

\begin{figure}[tbp]
\centerline{\psfig{figure=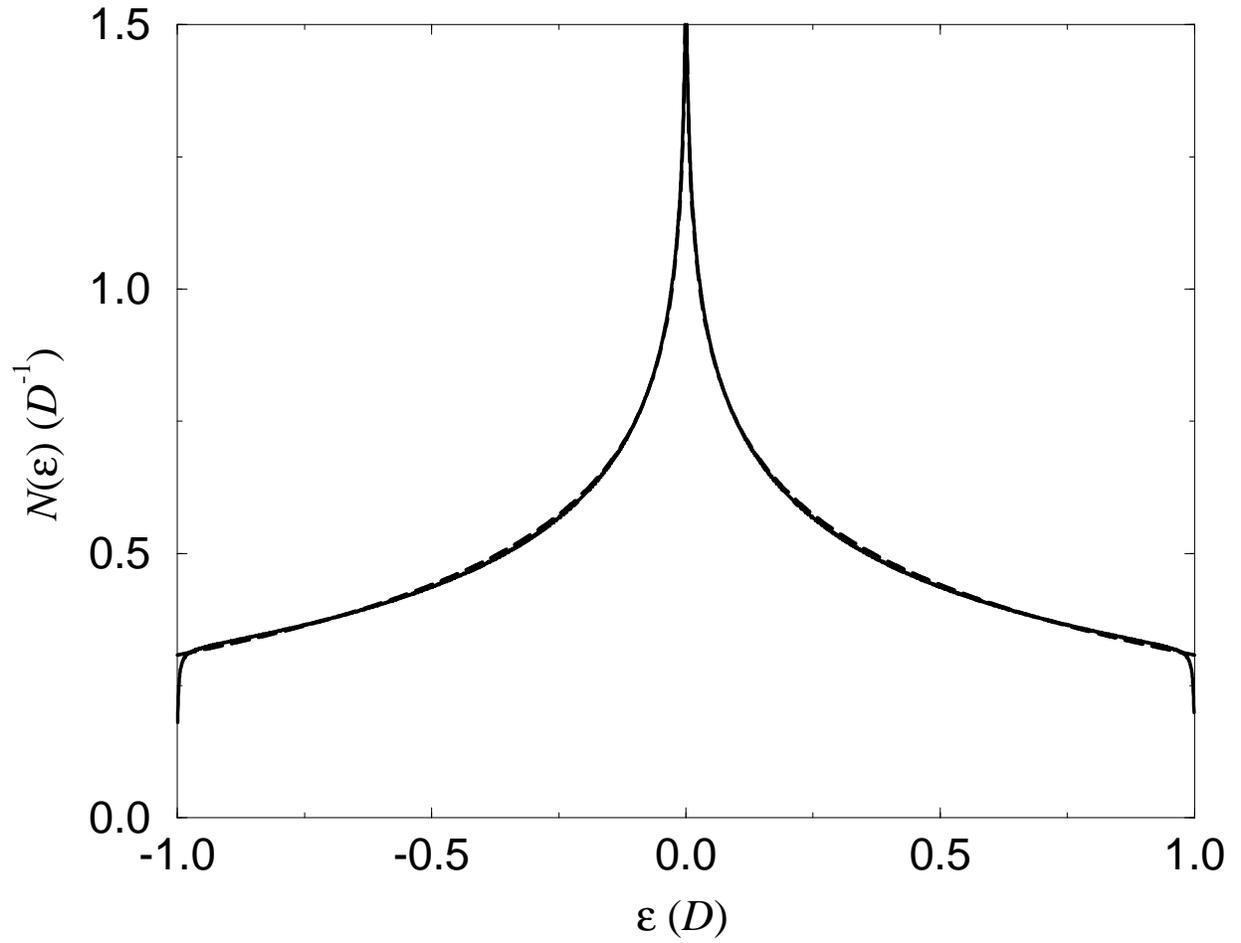,height=14.0cm,angle=270}}
\vskip1.0cm
\caption{The density of states N($\varepsilon$) for tight-binding electrons
on the square
lattice is shown as the solid curve. The energy scale is the
half-bandwidth $D=4t$. The best fit, the dashed line, is found
to be $N(\varepsilon)={0.31\over D}+{0.19\over D}\hbox{ln}\left|{D\over
\varepsilon}\right|$.}
\label{dos}
\end{figure}

In order to analytically solve the sums $I_{1-4}$, Eqs.\ (\ref{in}) and 
(\ref{i3}), we make the standard transformation to an integration
over energies, making use of the single-particle density of states (DOS) 

\begin{eqnarray}
N(\varepsilon)&\equiv&\sum_{\bf k}\delta(\varepsilon
-\varepsilon_k)\nonumber \\
&\approx&\cases{N(0)+N(1)\hbox{ln}\left|{D\over\varepsilon}\right|,& if 
$|\varepsilon|\le D$;\cr
0,&otherwise,\cr}
\end{eqnarray}

\noindent where $D=4t$ is the half-bandwidth 
and $\varepsilon_k\equiv\xi_k+\mu$. The DOS is approximated by a
constant plus a term reflecting the van Hove singularity at half-filling 
($\varepsilon=0$), as is shown in Fig.\ \ref{dos}. The best fit is 
obtained when $N(0)=0.31/D$
(the DOS for free electrons in two dimensions is $1/\pi D$) 
and $N(1)=0.19/D$ (note that the DOS at half-filling 
is approximately\cite{micnas} $(2/\pi^2D)\hbox{ln}|D/\varepsilon|$). 
Using the relation
$a_k=-2(\xi+\mu)/D$, the sums $I_{0-2}$ can be solved to yield:

\begin{equation}
I_0(T_s)\approx2N'\hbox{ln}\left({T^*\over T_s}\right)-{2D\over\mu}N(1);
\label{i0}
\end{equation}

\begin{equation}
I_1(T_s)\approx-{4\mu\over D}N'\left[\hbox{ln}\left({T^*\over T_s}\right)
-1\right]+{4\mu\over D}\left(1+{D\over\mu}\right)N(1);
\label{i1}
\end{equation}

\begin{equation}
I_2(T_s)\approx{8\mu^2\over D^2}N'\hbox{ln}\left({T^*
\over T_s}\right)+{4(D^2-3\mu^2)\over D^2}N(0)
+{2\over D^2}\bigg[-4\mu D+6\mu^2\hbox{ln}\left|{\mu\over D}\right|-5\mu^2
+D^2\bigg]N(1),
\end{equation}

\noindent where

\begin{equation}
N'=N(0)-N(1)\hbox{ln}\left|{\mu\over D}\right|,
\end{equation}

\noindent and

\begin{equation}
T^*\equiv{2e^{\gamma}\sqrt{D^2-\mu^2}\over\pi}.
\end{equation}

\noindent For all terms proportional to 
$N(1)$, we assume that $T_c\ll D$ (weak-coupling), so that

\begin{equation}
\hbox{tanh}\left({\varepsilon-\mu\over 2T}\right)\approx\cases{-1,
\varepsilon<\mu;\cr
+1,\varepsilon>\mu.\cr}
\end{equation}

\noindent Note that in this lattice model, all interactions are instantaneous
and therefore the bandwidth is the only possible energy cutoff. Since $b_k$ 
cannot be written exactly in terms of $\xi_k$, some simplifying
assumption must be made in order to evaluate $I_3$. At low densities (the
continuum limit of the lattice model), $\xi_k\approx tk^2-D-\mu$ and thus
$b_k\approx -2\hbox{cos}2\theta(\xi+D+\mu)/D$. Then,

\begin{eqnarray}
I_3(T_d)&\approx&{4(\mu+D)^2\over D^2}N'\hbox{ln}\left({T^*
\over T_d}\right)+{2\over D^2}\left(D^2-4\mu D-3\mu^2
\right)N(0)\nonumber \\
&+&{2\over D^2}\left[{-2D\over\mu}(\mu+D)^2+\mu(3\mu+4D)\hbox{ln}\left|
{\mu\over D}\right|+{D^2-8\mu D-5\mu^2\over 2}\right]N(1).
\end{eqnarray}

The equation for the s-wave transition temperature $T_s$ 
resulting from the application of Eq.\ (\ref{ts_cond}) is, for $V_0=0$:

\begin{equation}
T_s=T^*\hbox{exp}\left\{{-D^2-V_1\left[
6\mu^2N'-2D^2N(0)-\left(D^2-4\mu D-5\mu^2\right)N(1)\right]\over 4\mu^2V_1N'}
\right\}.
\end{equation}

\noindent The corresponding equation for the d-wave transition temperature
$T_d$ from Eq.\ (\ref{td_cond}) is:

\begin{equation}
T_d=T^*\hbox{exp}\left\{{-D^2-V_1\left[
\mu(4D+3\mu)N'-D^2N(0)+{1\over 2\mu}\left(4D^3+7D^2\mu+12D\mu^2+5\mu^3\right)
N(1)\right]\over 2(\mu+D)^2V_1N'}\right\}.
\end{equation}

\begin{figure}[tbp]
\centerline{\psfig{figure=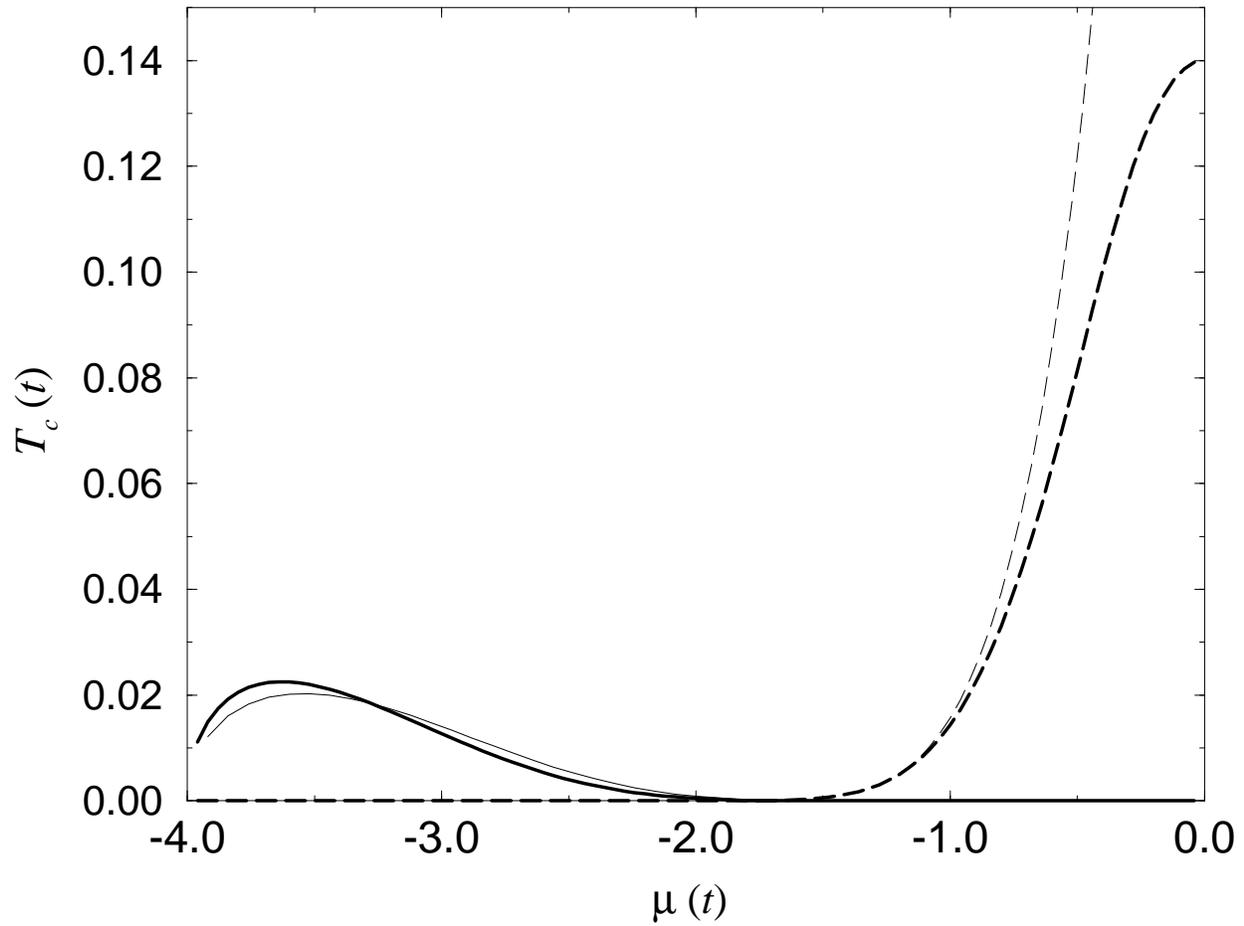,height=14.0cm,angle=270}}
\vskip1.0cm
\caption{The transition temperatures $T_c$ are given as functions of chemical
potential $\mu$ for nearest-neighbour attraction $V_1=t$ and $V_0=0$. The
results for $T_s$ and $T_d$ are
given by solid and dashed lines, respectively. The analytical results (lighter
lines) are found to compare well with the numerical results (darker lines) in
the applicable low-density regime.}
\label{tc1}
\end{figure}

\begin{figure}[tbp]
\centerline{\psfig{figure=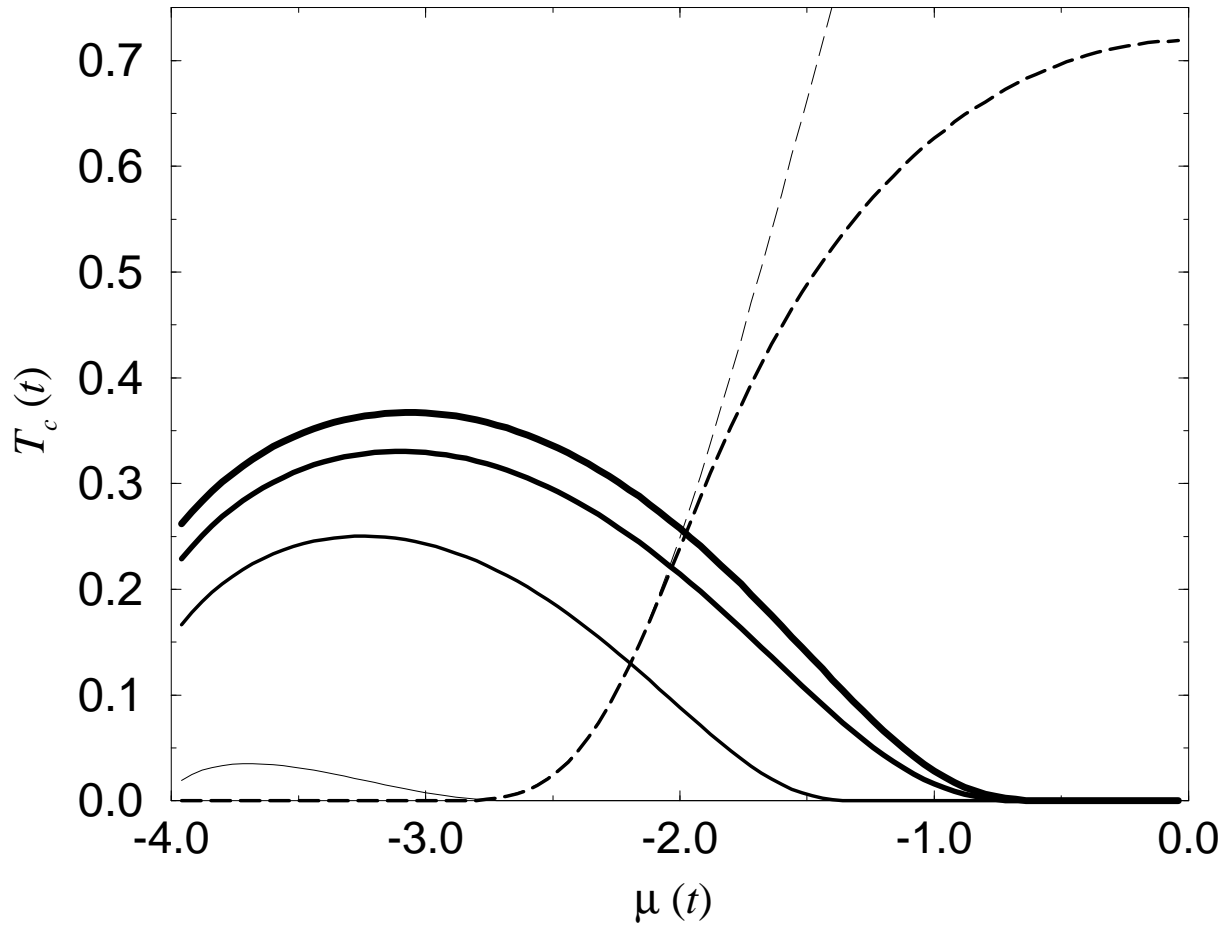,height=14.0cm,angle=270}}
\vskip1.0cm
\caption{The transition temperatures $T_c$ are given as functions of chemical
potential $\mu$ for $V_1=3t$. The solid lines correspond to the numerical
evaluation of $T_s$ for on-site repulsion $V_0=0$, $2.5t$, $3.5t$,
and $\sim\infty$; linewidth decreases with increased $V_0$. The d-wave
transition temperature $T_d$ (dashed lines) is unaffected by changes in $V_0$.
Note that the analytical (lighter dashed line) and numerical (darker dashed
line) results for $T_d$ still agree closely at low densities in this
intermediate-coupling regime. The correspondence between the numerical and
analytical results for $T_s$ (not shown) continues for intermediate coupling
but is found to improve with increased $V_0$, or decreased $T_s$.}
\label{tc3}
\end{figure}

\noindent The transition temperatures for the cases $V_1=t$ and
$V_1=3t$ are shown as
functions of chemical potential in Figs.\ \ref{tc1} and \ref{tc3} respectively. 
Near the bottom of the tight-binding band ($\mu\approx -D$), an
s-wave transition is strongly favoured for any $V_0$ and $V_1$, whereas a 
d-wave transition is favoured near half-filling ($\mu\approx 0$). The
value of the chemical potential at which the preferred symmetry of the dominant
gap function changes is extremely sensitive to the strengths of the 
respective interactions. As the on-site repulsion is increased, the magnitude 
of $T_s$ is suppressed while $T_d$ is unaffected.\cite{micnas} As a result, 
d-wave superconductivity is favoured for virtually all densities
for sufficiently large $V_0$. It should be noted that the magnitude of the 
subcritical transition temperature, associated with the subdominant gap
function, is in fact further decreased due to the presence of the dominant 
gap function. The relevant equation for the subcritical transition
temperature, one of Eqs.\ (\ref{delta0_eh})-(\ref{deltad_eh}), should have
$\xi_k$ replaced by $\sqrt{\xi_k^2+{\Delta_k^{\rm DOM}}^2}$, 
where $\Delta_k^{\rm DOM}$ is the magnitude of the dominant gap funcion.
Thus, the 
subcritical transition temperature, which is of dubious relevance in any case,
can be taken to be identically zero without loss of generality. 

\section{Calculation of the Ginzburg-Landau coefficients}

\subsection{Extended Hubbard Model}

For sufficiently large systems, the lattice sums in Eq.\ (\ref{gaps_eh})
can be transformed into k-space integrals. Neglecting fourth-order derivatives
and making use of the expressions for 
the s-wave (\ref{def_s}) and d-wave (\ref{def_d}) gap functions as well as the 
normal-state Green function (\ref{green}), the following three gap
equations for the EH model are obtained:

\begin{eqnarray}
\Delta^*_{\circ}&=&4TV_0\sum_{\omega_n}\int{d^2k\over(2\pi)^2}\bigg\{
{s^*({\bf r},{\bf k})\over\omega_n^2+\xi_k^2}-{1\over(\omega_n^2+\xi_k^2)^2}
\left[16|s({\bf r},{\bf k})|^2s^*({\bf r},{\bf k})
+32|d({\bf r},{\bf k})|^2s^*({\bf r},{\bf k})+16{d^*}^2({\bf r},{\bf k})
s({\bf r},{\bf k})\right]\nonumber \\
&-&\int d^2z{z^2\over 2}\int{d^2k'\over(2\pi)^2}{e^{i({\bf k}+{\bf k}')\cdot
{\bf z}}\over(-i\omega_n-\xi_{k'})(i\omega_n-\xi_k)}\left(\hbox{cos}^2\theta
\Pi_x^2+\hbox{sin}^2\theta\Pi_y^2\right)\left[s^*({\bf r},{\bf k})
+d^*({\bf r},{\bf k})\right]\bigg\}; \label{delta0b_eh} \\
\Delta^*_s&=&2TV_1\sum_{\omega_n}\int{d^2k\over(2\pi)^2}a_k\bigg\{
\hbox{as above}\bigg\} \label{deltasb_eh} \\
\Delta^*_d&=&2TV_1\sum_{\omega_n}\int{d^2k\over(2\pi)^2}b_k\bigg\{
{d^*({\bf r},{\bf k})\over\omega_n^2+\xi_k^2}-{1\over(\omega_n^2+\xi_k^2)^2}
\left[16|d({\bf r},{\bf k})|^2d^*({\bf r},{\bf k})+32|s({\bf r},{\bf k})|^2
d^*({\bf r},{\bf k})+16{s^*}^2({\bf r},{\bf k})d({\bf r},
{\bf k})\right]\nonumber \\
&-&\int d^2z{z^2\over 2}\int{d^2k'\over(2\pi)^2}{e^{i({\bf k}+{\bf k}')\cdot
{\bf z}}\over(-i\omega_n-\xi_{k'})(i\omega_n-\xi_k)}\left(\hbox{cos}^2\theta
\Pi_x^2+\hbox{sin}^2\theta\Pi_y^2\right)\left[s^*({\bf r},{\bf k})
+d^*({\bf r},{\bf k})\right]\bigg\}\label{deltadb_eh}, 
\end{eqnarray}

\noindent where 
$s({\bf r},{\bf k})\equiv{a_k\over 2}\Delta_s({\bf r})-{1\over 4}\Delta_{\circ}
({\bf r})$, $d({\bf r},{\bf k})\equiv{b_k\over 2}\Delta_d({\bf r})$, 
$\theta$ is the angle between {\bf r} and {\bf z} ($=m\hat{x}+n\hat{y}$), and
the continuum limit of (\ref{def_grad})

\begin{equation}
\Pi_z\equiv -i{\partial\over\partial z}+{4\pi\over\phi_0}A_z({\bf r})
\end{equation}

\noindent acts only on the center of mass coordinate {\bf r}.
Eqs.\ (\ref{delta0b_eh})-(\ref{deltadb_eh}) can be simplified by making use
of the fact that $\Delta_{\circ}({\bf r})=\epsilon\Delta_s({\bf r})$, where
$\epsilon$ depends on temperature and chemical potential. This relation follows
from the observation that (\ref{delta0b_eh}) and (\ref{deltasb_eh}) are not
linearly independent, since $\Delta_{\circ}({\bf r})$ and $\Delta_s({\bf r})$
have the same symmetry. Near $T_c$ the magnitude of the s-wave component is 
small; from (\ref{delta0_eh}) we infer $\epsilon\approx V_0I_1/(1+V_0I_0/2)$,
where $I_0$ and $I_1$ are defined in Eq.\ (\ref{in}) and are 
evaluated in the weak-coupling and low-density limit in Eqs.\ (\ref{i0}) 
and (\ref{i1}). The integrals involving gradients of the gap 
functions can be greatly simplified by means of the identity:

\begin{equation}
\int d^2z\; x^2 G^{\circ}({\bf z},-\omega_n)G^{\circ}({\bf z}
+\vec{\alpha},\omega_n)
\equiv\int{d^2k\over(2\pi)^2}e^{-i{\bf k}\cdot\vec{\alpha}}\Big|
\nabla_{k_x}G^{\circ}
({\bf k},\omega_n)\Big|^2,
\label{trick}
\end{equation}

\noindent and similarly for integrals involving $y^2$. Thus,

\begin{eqnarray}
{1\over 4}\left(1-{V_1\epsilon^2\over 4V_0}\right)\Delta_s^*({\bf r})
&=&TV_1\sum_{\omega_n}\int{d^2k\over(2\pi)^2}{a_k'\over\omega_n^2+\xi_k^2}
\bigg\{a_k'\Delta_s^*({\bf r})\nonumber \\
&-&{1\over\omega_n^2+\xi_k^2}\Big[16{a_k'}^3
|\Delta_s({\bf r})|^2\Delta_s^*({\bf r})+8a_k'b_k^2|\Delta_d({\bf r})|^2
\Delta_s^*({\bf r})+4a_k'b_k^2{\Delta_d^*}^2({\bf r})\Delta_s({\bf r})
\nonumber \\
&+&{D^2\over 16}\left(\hbox{sin}^2k_x\Pi_x^2+\hbox{sin}^2k_y\Pi_y^2\right)
\left[2a_k'\Delta_s^*({\bf r})+b_k\Delta_d^*({\bf r})\right]\Big]\bigg\},
\label{s_eh}
\end{eqnarray}

\begin{eqnarray}
{1\over 4}\Delta_d^*({\bf r})
&=&TV_1\sum_{\omega_n}\int{d^2k\over(2\pi)^2}{b_k\over\omega_n^2+\xi_k^2}
\bigg\{{1\over 4}b_k\Delta_d^*({\bf r})\nonumber \\
&-&{1\over\omega_n^2+\xi_k^2}\Big[b_k^3
|\Delta_d({\bf r})|^2\Delta_d^*({\bf r})+8{a_k'}^2b_k|\Delta_s({\bf r})|^2
\Delta_d^*({\bf r})+4{a_k'}^2b_k{\Delta_s^*}^2({\bf r})\Delta_d({\bf r})
\nonumber \\
&+&{D^2\over 32}\left(\hbox{sin}^2k_x\Pi_x^2+\hbox{sin}^2k_y\Pi_y^2\right)
\left[2a_k'\Delta_s^*({\bf r})+b_k\Delta_d^*({\bf r})\right]\Big]\bigg\},
\label{d_eh}
\end{eqnarray}

\noindent where $a_k'\equiv {a_k\over 2}-{\epsilon\over 4}$.

The appropriate free energy for a d-wave (or extended s-wave) superconductor 
on a tetragonal lattice is:\cite{joynt}

\begin{eqnarray}
F_s&=&F_n+\alpha_s|\Delta_s({\bf r})|^2+\alpha_d|\Delta_d({\bf r})|^2
+\beta_1|\Delta_s({\bf r})|^4+\beta_2|\Delta_d({\bf r})|^4\nonumber \\
&+&\beta_3|\Delta_s({\bf r})|^2|\Delta_d({\bf r})|^2+\beta_4\left(
{\Delta_s^*({\bf r})}^2\Delta_d^2({\bf r})+{\Delta_d^*({\bf r})}^2
\Delta_s^2({\bf r})\right)\nonumber \\
&+&\gamma_s|\vec{\Pi}\Delta_s({\bf r})|^2+\gamma_d|\vec{\Pi}\Delta_d({\bf r})|^2
+\gamma_{\nu}\Big[\big(\Pi_y\Delta_s({\bf r})\big)^*\big(\Pi_y\Delta_d({\bf r})
\big)-\big(\Pi_x\Delta_s({\bf r})\big)^*\big(\Pi_x\Delta_d({\bf r})\big)
+c.c.\Big]\nonumber \\
&+&F_1^s+F_1^d+{h^2\over 8\pi},
\label{gl}
\end{eqnarray}

\noindent where higher-order terms

\begin{equation}
F_1^s=\eta_s|\Delta_s({\bf r})\vec{\Pi}\Delta_s({\bf r})|^2
+\gamma_{s+}|\vec{\Pi}^2\Delta_s
({\bf r})|^2+\gamma_{s-}|(\Pi_y^2-\Pi_x^2)\Delta_s({\bf r})|^2,
\label{f1s}
\end{equation}

\begin{equation}
F_1^d=\eta_d|\Delta_d({\bf r})\vec{\Pi}\Delta_d({\bf r})|^2
+\gamma_{d+}|\vec{\Pi}^2\Delta_d
({\bf r})|^2+\gamma_{d-}|(\Pi_y^2-\Pi_x^2)\Delta_d({\bf r})|^2
\label{f1d}
\end{equation}

\noindent will be discussed
in detail in Section V. In the present work, the s-wave and d-wave order 
parameters are respectively the gap functions
$\Delta_s({\bf r})$ and $\Delta_d({\bf r})$. The GL free energy (\ref{gl})
implies an s-wave (d-wave) component is induced whenever there
exist spatial variations of the dominant d-wave (s-wave) order parameter. The
magnitude of the `subdominant' order parameter is evidently proportional to
the coefficient $\gamma_{\nu}$ of the mixed gradient term.
Minimizing and comparing to Eqs.\ (\ref{s_eh}) and (\ref{d_eh}) immediately 
yields: 

\begin{equation}
\alpha_s={1\over V_1}-{\epsilon^2\over 4V_0}-4T\sum_{\omega_n}\int{d^2k
\over(2\pi)^2}{{a_k'}^2\over\omega_n^2+\xi_k^2}\quad ;\quad
\alpha_d={1\over V_1}-T\sum_{\omega_n}\int{d^2k\over(2\pi)^2}
{b_k^2\over\omega_n^2+\xi_k^2},\label{alpha_eh}
\end{equation}

\begin{equation}
\{\beta_1\;,\;\beta_2\;,\;\beta_3\;,\;\beta_4\}=T\sum_{\omega_n}\int
{d^2k\over(2\pi)^2}{\{32{a_k'}^4\;,\;2b_k^4\;,\;32{a_k'}^2b_k^2\;,\;8{a_k'}^2
b_k^2\}\over(\omega_n^2+\xi_k^2)^2},\label{beta_eh}
\end{equation}

\begin{equation}
\{\gamma_s\;,\;\gamma_{\nu}\;,\;\gamma_d\}={D^2\over 8}T\sum_{\omega_n}\int
{d^2k\over(2\pi)^2}\hbox{sin}^2k_x{\{4{a_k'}^2\;,\;-2a_k'b_k\;,\;b_k^2\}\over
(\omega_n^2+\xi_k^2)^2}.\label{gamma_eh}
\end{equation}

\noindent The current density in the $xy$ plane can be obtained either by 
evaluating (\ref{cur_eh}) in the continuum limit or by minimizing the free 
energy (\ref{gl}) with respect to the vector potential:

\begin{eqnarray}
{\bf j}&=&\delta{4\pi c\over\phi_0}\bigg[\gamma_d\Delta_d^*({\bf r})\vec{\Pi}
\Delta_d({\bf r})+\gamma_s\Delta_s^*({\bf r})\vec{\Pi}\Delta_s({\bf r})
-\hat{x}\gamma_{\nu}\Big(\Delta_s^*({\bf r})\Pi_x\Delta_d({\bf r})
+\Delta_d^*({\bf r})\Pi_x\Delta_s({\bf r})\Big)\nonumber \\
& &\qquad +\hat{y}\gamma_{\nu}\Big(\Delta_s^*({\bf r})\Pi_y\Delta_d({\bf r})
+\Delta_d^*({\bf r})\Pi_y\Delta_s({\bf r})\Big)+c.c.\bigg],
\label{current}
\end{eqnarray}

\noindent where the factor $\delta=$(layer thickness/layer spacing) is 
introduced in order to model the layered structure of the cuprate 
superconductors. In the present two-dimensional model,
$\delta\rightarrow 0$, reflecting the inability of the system to 
sustain screening currents. Eqs.\ (\ref{alpha_eh})-(\ref{current}) are subject
to the following boundary conditions:

\begin{eqnarray}
& &{\bf n}\cdot\big[\gamma_d\vec{\Pi}\Delta_d({\bf r})
+\gamma_{\nu}\left(\hat{y}\Pi_y-\hat{x}\Pi_x\right)\Delta_s({\bf r})\big]=0,\\
& &{\bf n}\cdot\big[\gamma_s\vec{\Pi}\Delta_s({\bf r})
+\gamma_{\nu}\left(\hat{y}\Pi_y-\hat{x}\Pi_x\right)\Delta_d({\bf r})\big]=0,
\end{eqnarray}

\noindent where {\bf n} is the unit vector normal to the surface of the 
superconductor. 

The coefficients of the GL free energy can be calculated analytically
in the same low-density and weak-coupling limit employed earlier by setting 
$a_k\approx -2\mu/D$, 
$b_k\approx -2(\mu+D)\hbox{cos}2\theta/D$ and $\hbox{sin}^2k_x\approx 4(\mu+D)
\hbox{cos}^2\theta/D$. Making use of

\begin{equation}
T\sum_{\omega_n}\int{d^2k\over(2\pi)^2}{1\over(\omega_n^2+\xi_k^2)^2}
\approx{7\zeta(3)N'\over 8\pi^2T^2},
\end{equation}

\noindent we immediately obtain:

\begin{equation}
\alpha_s= -{1\over V_1}\left({\mu\over D}+{\epsilon\over 4}
\right)^2\left(1-{T\over T_s}\right)\quad ;\quad\alpha_d= -{1\over V_1}
\left({\mu+D\over D}\right)^2\left(1-{T\over T_d}\right),\label{alpha_eh_ans}
\end{equation}

\begin{equation}
\{\beta_1\;,\;\beta_2\;,\;\beta_3\;,\;\beta_4\}=4\gamma\left\{\left({\mu\over D}
+{\epsilon\over 4}\right)^4,\;{3\over 8}\left({\mu+D\over D}\right)^4,\;
2\left({\mu\over D}+{\epsilon\over 4}\right)^2\left({\mu+D\over D}\right)^2,
\;{1\over 2}\left({\mu\over D}+{\epsilon\over 4}\right)^2\left({\mu+D\over D}
\right)^2\right\},\label{beta_eh_ans}
\end{equation}

\begin{equation}
\{\gamma_s\;,\;\gamma_{\nu}\;,\;\gamma_d\}=\gamma{v_F^2\over 16}\left\{2
\left({\mu\over D}+{\epsilon\over 4}\right)^2,\;-\left({\mu\over D}+{\epsilon
\over 4}\right)\left({\mu+D\over D}\right),\;\left({\mu+D\over D}\right)^2
\right\}\label{gamma_eh_ans}
\end{equation}

\noindent where 

\begin{equation}
\gamma={7\zeta(3)N'\over\pi^2T^2},
\end{equation}

\noindent and $v_F^2=D^2k_F^2/4\approx D(\mu+D)$. From (\ref{i0}) 
and (\ref{i1}) we obtain:

\begin{equation}
\epsilon= -{4\mu\over D}{V_0N'\hbox{ln}\left({T^*\over T}\right)-V_0N'
\over V_0N'\hbox{ln}\left({T^*\over T}\right)+1}.
\end{equation}

\noindent The  expressions for $\alpha_s$ and $\alpha_d$ are only valid 
near their respective critical temperatures.
Since the subcritical transition temperature is much lower than the dominant
transition temperature, the corresponding coefficient for the subdominant order 
parameter can be assumed approximately $1/V_1$ for all values of $T<T_c$ (i.e.\
the contribution of the appropriate integral in either of (\ref{alpha_eh}) will
be negligible).  Note that in the limit $V_1=0$, $V_0\rightarrow -V_0$,
$\Delta_s$ and $\Delta_d$ vanish; the relevant gap equation is therefore Eq.\ 
(\ref{delta0b_eh}) with $s({\bf r},{\bf k})=\Delta_0({\bf r})$ 
and $d({\bf r},{\bf k})=0$. Then Eq.\ (\ref{gl}) becomes the two-dimensional 
continuum BCS GL free energy:

\begin{equation}
F_s^{\rm BCS}=F_{n}-{1\over V_0}\left(1-{T\over T_s}\right)|\Delta_{\circ}
({\bf r})|^2
+{7\zeta(3)N(0)\over 16\pi^2T^2}|\Delta_{\circ}({\bf r})|^4
+{7\zeta(3)N(0)\over 32\pi^2T^2}v_F^2\big|\vec{\Pi}\Delta_{\circ}({\bf r})
\big|^2+{h^2\over 8\pi}.
\end{equation}

The GL coefficients (\ref{alpha_eh_ans})-(\ref{gamma_eh_ans}) of the free 
energy (\ref{gl}) imply that at the bottom of the band ($\mu=-D$) the d-wave
component vanishes, yielding a pure s-wave superconductor, while at 
half-filling ($\mu=0$) only the d-wave component remains (with the caveat that 
these analytical results become decreasingly valid near half-filling). At 
densities intermediate between these
two extremes all the GL coefficients are non-zero, leading to coexisting d-wave 
and s-wave order paramaters for any temperature $T<T_c$ and finite external 
magnetic field. For type-II superconductors described by (\ref{gl}) in fields
just above $H_{c1}$, it has been found\cite{berlinsk,ren,franz2} that the 
subdominant order parameter is nucleated in the vicinity of a magnetic vortex 
core whenever the coefficient of the mixed-gradient term $\gamma_{\nu}$ is
non-zero. Moving in the $x=y$ direction from the 
center of the vortex, the induced subdominant order parameter reaches a
maximum, then decreases algebraically; there exist extra nodes 
in the $x=0$ or $y=0$ directions. The maximum amplitude 
attained by the subdominant order parameter nucleated in the vortex core is 
given by:\cite{franz2}

\begin{equation}
{\Delta_s(\Delta_d)^{\rm max}\over\Delta_d(\Delta_s)^{\rm bulk}}
\approx{3\over 16}{\gamma_{\nu}\over
\gamma_{d(s)}}{|\alpha_{d(s)}|\over\alpha_{s(d)}}\approx{3\over 16}
\left(1-{T\over T_{d(s)}}\right)
{\gamma_{\nu}\over\gamma_{d(s)}},\qquad T_{d(s)}>T_{s(d)},\label{maxsub}
\end{equation}

\noindent where $\Delta_s^{\rm max}$ is the maximum value of the induced 
subdominant 
s-wave order parameter compared with the magnitude of the critical d-wave order
parameter $\Delta_d^{\rm bulk}$ far from the vortex core, and vice versa.
The estimates (\ref{maxsub}) are reliable as long 
as $\Delta_s(\Delta_d)^{\rm max}<\Delta_d(\Delta_s)^{\rm bulk}/4$, which is 
always the case sufficiently
near $T_c$: since $\Delta_d(\Delta_s)^{\rm bulk}\sim\sqrt{(1-T/T_c)}$, the 
above equations indicate that 

\begin{equation} 
\Delta_s(\Delta_d)^{\rm max}\sim(1-T/T_c)^{3/2}.
\label{subdom}
\end{equation}

\noindent The subdominant
order parameter must decay more rapidly near the transition temperature
since it is induced through spatial variations of the critical order parameter.

From (\ref{gamma_eh_ans}), we have:

\begin{eqnarray}
{\gamma_{\nu}\over\gamma_d}&\approx&\left|{\mu+D\epsilon/4\over\mu+D}\right|,
\qquad T_d>T_s.\label{ratio1} \\
{\gamma_{\nu}\over\gamma_s}&\approx&{1\over 2}\left|{\mu+D\over\mu+D\epsilon/4}
\right|,\qquad T_s>T_d,\label{ratio2}
\end{eqnarray}

\begin{figure}[tbp]
\centerline{\psfig{figure=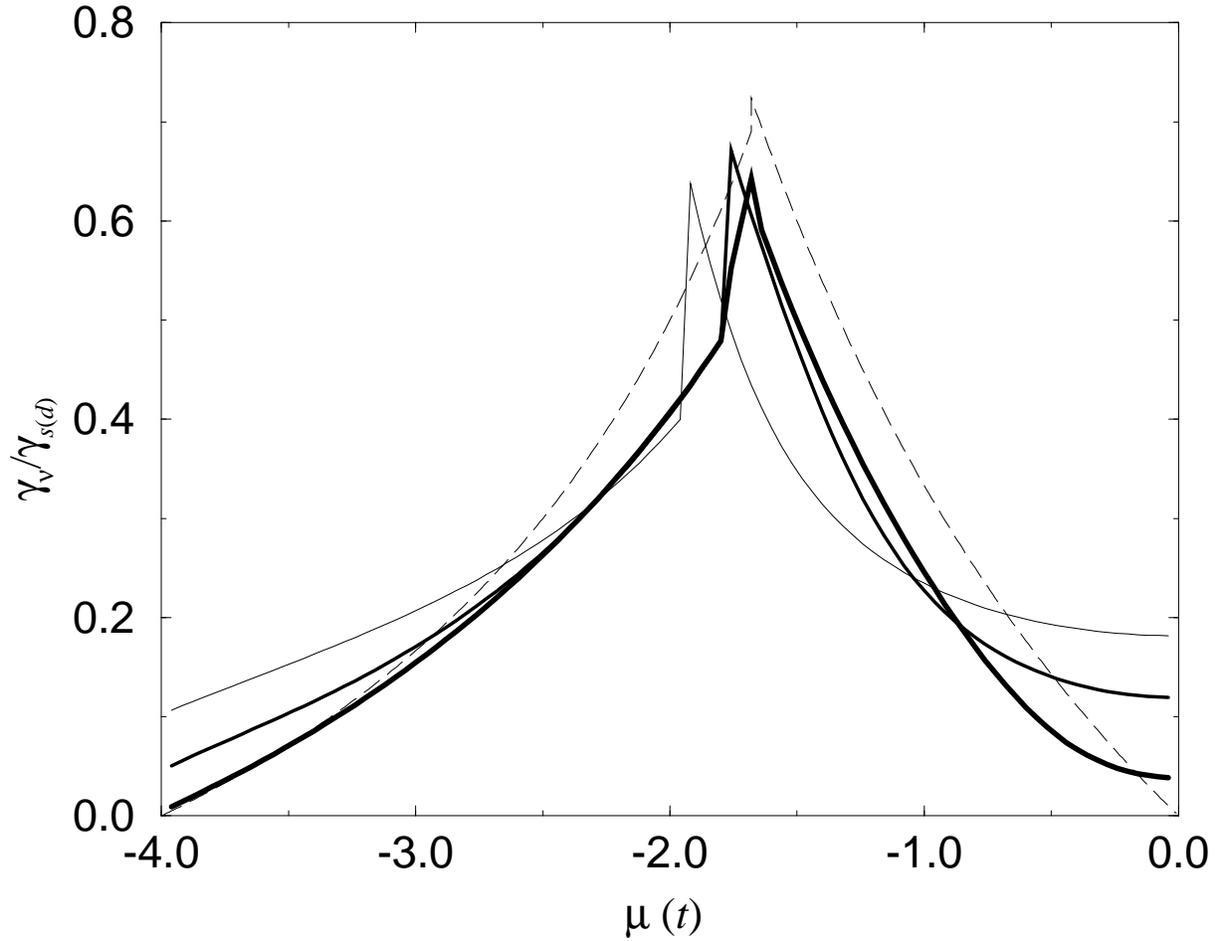,height=14.0cm,angle=270}}
\vskip1.0cm
\caption{The ratio of the mixed gradient coefficient $\gamma_{\nu}$ to the
ordinary gradient coefficient of the dominant order parameter $\gamma_s$ or
$\gamma_d$ is given as a function of chemical potential $\mu$. The solid and
dashed lines correspond to numerical and analytical results respectively. The
solid lines become progressively darker as $V_1$ is decreased from $3t$
to $t$ in increments of $t$; $V_0$ is taken to be zero. The discontinuity
reflects the transition from an s-wave to d-wave bulk superconductor; the
dashed line assumes that this occurs for $\mu=-1.65t$. All values are
determined at the appropriate $T_c$.}
\label{grads_eh}
\end{figure}

\begin{figure}[tbp]
\centerline{\psfig{figure=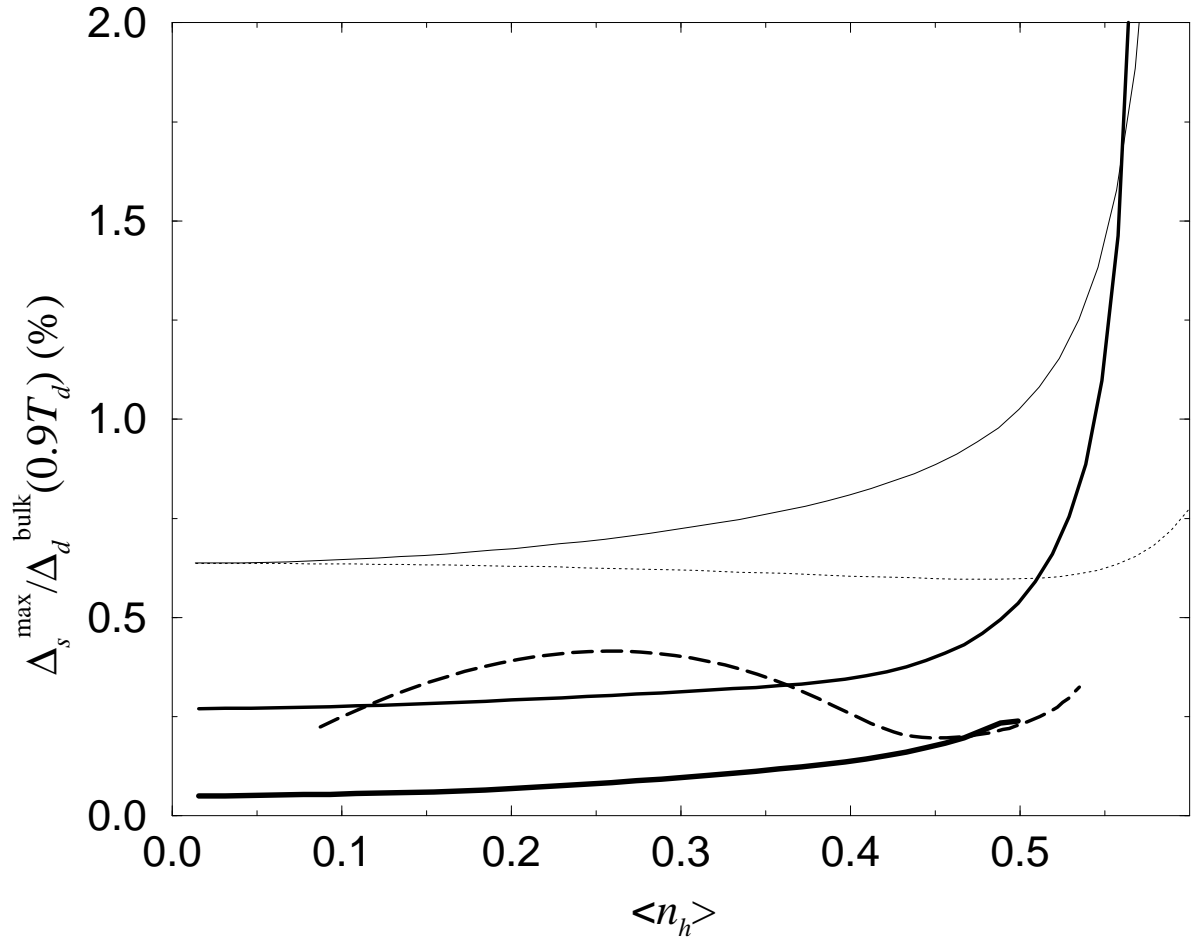,height=14.0cm,angle=270}}
\vskip1.0cm
\caption{The relative magnitude of the maximum s-wave component induced in the
vortex core $s_{\rm max}/d_{\rm bulk}$ is calculated numerically and given as a
function of hole density $\langle n_h\rangle$ for temperatures $T=0.9T_d$. The
dashed line, corresponding to results for the AvH model,
indicates that the induced s-wave component is largest near optimal doping
($\langle n_h\rangle\sim 20\%$). Results for the EH model are given for
comparison, where in the hole representation $\langle n_h\rangle =0$
corresponds to $\mu=0$; solid lines (in
order of decreasing boldness) correspond to $V_1=t,2t,3t$ with $V_0=0$, while
the dotted line is for $V_1=3t,V_0=4t$.}
\label{max_all}
\end{figure}

\noindent The gradient coefficient ratios (\ref{ratio1}) and (\ref{ratio2}),
which govern the magnitude of the subdominant order, are compared 
with the appropriate numerical results in Fig.\ \ref{grads_eh} for the special
case $V_0=0$. The analytical
results capture the essential physics in their regime of applicability, i.e.\
at low densities and weak to intermediate coupling (better quantitative 
agreement at intermediate coupling can be obtained by including in $a_k$,
$b_k$, etc.\ terms higher order in $\xi_k$). Near half-filling (zero-filling),
where d-wave (s-wave) superconductivity is most stable, $\gamma_{\nu}/\gamma_d$ 
($\gamma_{\nu}/\gamma_s$) grows with increased coupling. As a result,
a significant component of the subdominant order parameter would only exist for
stronger coupling in these two density regimes. It should be kept in mind, 
however, that as the on-site repulsion $V_0$ (and therefore $\epsilon$) 
increases from zero, the 
magnitude of the s-wave component is suppressed for all densities; a 
d-wave subcomponent would be thereby enhanced in a bulk s-wave superconductor.
The magnitude of the subdominant order parameter can be significantly larger
even for weak coupling at intermediate densities, however; not only is the 
gradient coefficient ratio enhanced,
but also the ratio $|\alpha_{\rm dom}|/\alpha_{\rm sub}$ 
since $\alpha_{\rm sub}\rightarrow 0$ as the subdominant superconducting 
instability is approached. (Fig.\ \ref{max_all} in the following section 
provides further quantitative details).

There are in general two characteristic length scales $\xi_s$ and $\xi_d$
governing spatial variations of the s-wave and d-wave order parameters in the 
vicinity of a vortex core, respectively. Near $T_c$, however, the
induced subdominant order parameter is negligible compared to the critical 
order parameter by (\ref{subdom}). Keeping only terms in the free 
energy (\ref{gl}) 
involving the dominant order parameter, the superconducting 
coherence lengths $\xi(T)$ and penetration depths $\lambda(T)$ near $T_c$ 
can be crudely estimated:

\begin{eqnarray}
\xi_{s(d)}(T)&=&\sqrt{{\gamma_{s(d)}\over|\alpha_{s(d)}|}}\label{cohere} \\
\lambda_{s(d)}(T)&=&\sqrt{{2\phi_0^2\beta_{1(2)}\over\delta(4\pi)^3\gamma_{s(d)}
|\alpha_{s(d)}|}}\label{lambda}
\end{eqnarray}

\noindent where the flux quantum $\phi_0=hc/e$ and both lengths are given in 
units of the lattice constant $a$. Inserting the relevant analytical 
expressions from Eqs.\ (\ref{alpha_eh_ans})-(\ref{gamma_eh_ans}), valid
in the low-density, weak-intermediate coupling limit, we obtain:

\begin{equation}
\xi_s(T)\approx\left({v_F\over 2\pi T}\right)\sqrt{{7\zeta(3)V_1N'\over 2
(1-T/T_s)}}\quad ;\quad
\xi_d(T)\approx\left({v_F\over 2\pi T}\right)\sqrt{{7\zeta(3)V_1N'\over 4
(1-T/T_d)}};
\end{equation}

\begin{equation}
\lambda_s(T)\approx\sqrt{{\phi_0^2V_1\over\pi^3\delta v_F^2(1-T/T_s)}}\quad ;
\quad \lambda_d(T)\approx\sqrt{{3\phi_0^2V_1\over 4\pi^3\delta v_F^2(1-T/T_d)}}.
\end{equation}

\begin{figure}[tbp]
\centerline{\psfig{figure=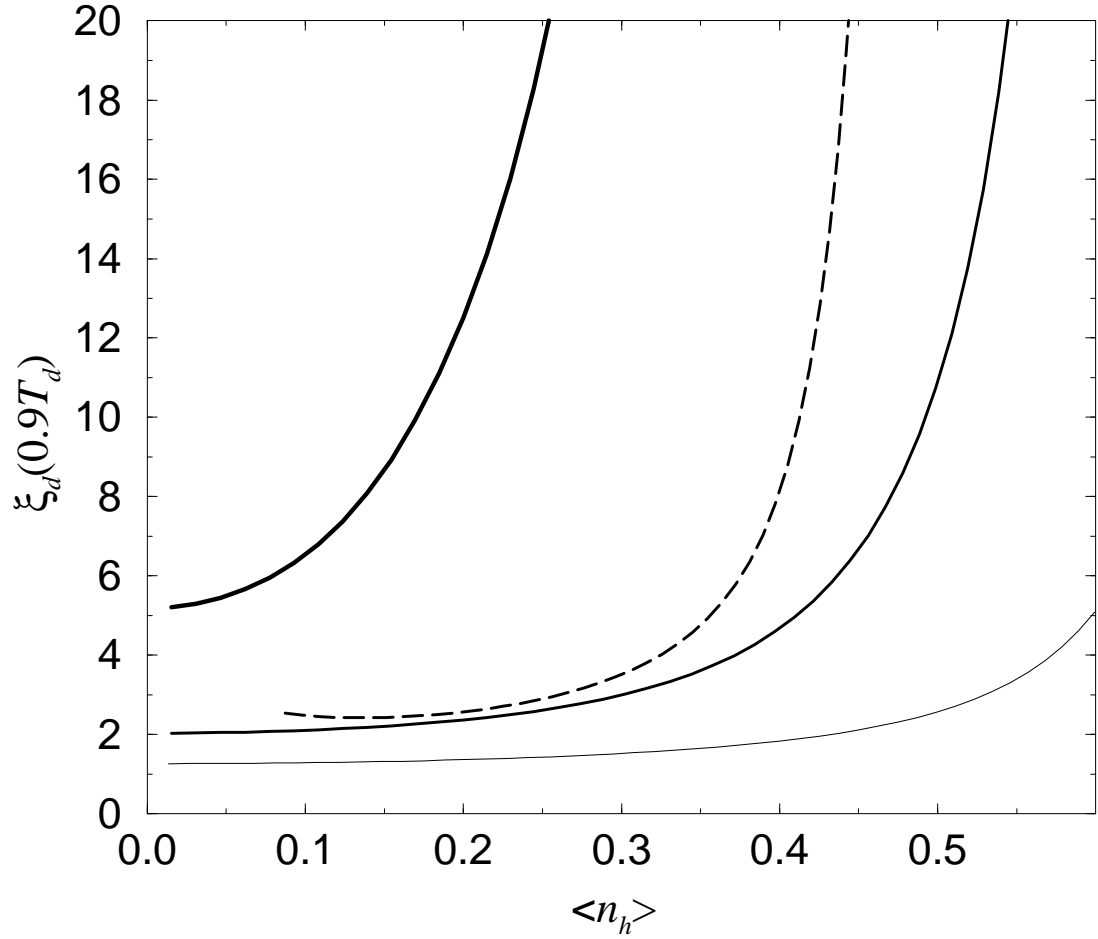,height=14.0cm,angle=270}}
\vskip1.0cm
\caption{The numerical d-wave Ginzburg-Landau coherence length $\xi_d(T)$,
measured in units of a lattice spacing, is shown for
$T=0.9T_d$ as a function of hole concentration $\langle n_h\rangle$. The
dashed line corresponds to the AvH model while the solid lines (in order of
decreasing boldness) correspond to the EH model for $V_1=t,2t,3t$ and $V_0=0$.}
\label{coh_all}
\end{figure}

\noindent Note that the penetration depths are effectively infinite 
for an isolated layer. The magnitudes of the coherence lengths can be estimated 
for parameters stabilizing either s-wave or d-wave superconductivity (depending
on the magnitude of $V_0$). For example, with $\mu=-2.0t$, $T_c\sim 0.2t$ 
($V_1\sim 3t$), and $T=0.9T_c$, we obtain $\xi_d\sim 6a$ and $\xi_s\sim 8a$ for
d-wave and s-wave superconductors respectively. 
The coherence lengths tend to become progressively shorter with increased 
electron density (Fig.\ \ref{coh_all} in the next section gives further 
details).

It is useful to compare the GL free energy (\ref{gl}) with that derived by
Ren {\em et al.}\cite{ren} Defining $s\equiv -\left({\mu\over D}
+{\epsilon\over 4}\right)\Delta_s$ and $d\equiv{\mu+D\over D}\Delta_d$ (we drop 
reference to the center of mass coordinate {\bf r} for convenience), we obtain:

\begin{eqnarray}
F_s&=&F_n-\left(1-{T\over T_s}\right)|s|^2-\left(1-{T\over T_d}\right)
|d|^2+4\gamma\Big\{|s|^4+{3\over 8}|d|^4+2|s|^2|d|^2+{1\over 2}\left(
{s^*}^2d^2+{d^*}^2s^2\right)\Big\} \nonumber \\
&+&\gamma{v_F^2\over 16}\Big\{2|\vec{\Pi}s|^2+|\vec{\Pi}d|^2
+\Big[\big(\Pi_ys\big)^*\big(\Pi_yd\big)-\big(\Pi_xs
\big)^*\big(\Pi_xd\big)+c.c.\Big]\Big\}+{h^2\over 8\pi}.
\label{gl_eh}
\end{eqnarray}

\noindent While the various coefficients appearing in the above free energy 
appear to be similar to those found by Ren {\em et al.}, one important 
distinction must be emphasized. The above free energy (\ref{gl_eh}) mixes a 
d-wave order parameter with an {\em extended} s-wave order parameter. The
analytical results suggest that only near half-filling ($\mu\rightarrow 0$) 
does the contribution of $\Delta_s$ to $s$ vanish. At these densities, however,
the approximations employed in the microscopic calculation above are not
strictly valid; indeed, the numerical results clearly indicate the existence of
a non-vanishing s-wave component for $T<T_d$ even in the absence of an
on-site interaction (see Fig.\ \ref{grads_eh}).
For any other density (except $\mu\rightarrow -D$, where the d-wave 
component vanishes) at temperatures below $T_c$, both $\Delta_s$ and $\Delta_d$
will be non-zero in the vortex core since both are generated by the same
nearest-neighbour pairing interaction $V_1$.
There is therefore no limit in which the present model reduces to that of 
Ren {\em et al.}, i.e.\ where an isotropic s-wave order parameter derived from
a repulsive on-site interaction is alone nucleated by spatial variations of a 
dominant d-wave order parameter. The technical difficulty
which necessitated the implementation of the Pad\'e approximation in 
Ref.\ \onlinecite{ren} was due to the indistinguishability of $\Delta_{\circ}$
and $\Delta_s$ in the continuum. This problem can always be avoided by 
starting with a lattice model which gives rise to a d-wave order parameter, 
then taking the appropriate continuum limit.

\subsection{Antiferromagnetic-van Hove Model}

Neglecting fourth-order derivatives, the equations for the gap functions in 
the thermodynamic limit of the AvH model from Eq.\ (\ref{gaps_avh}) are:

\begin{eqnarray}
\Delta^*_s&=&{TV\over 2}\sum_{\omega_n}\int{d^2k\over(2\pi)^2}\bigg\{
{c_k\Delta^*_s({\bf r})\over\omega_n^2+\xi_k^2}
-{1\over 4(\omega_n^2+\xi_k^2)^2}\left[c_k^2|\Delta_s({\bf r})|^2
\Delta^*_s({\bf r})
+2c_kd_k|\Delta_d({\bf r})|^2\Delta^*_s({\bf r})+c_kd_k{\Delta_d^*}^2
({\bf r})\Delta_s({\bf r})\right]\nonumber \\
&-&\int d^2z{z^2\over 2}\int{d^2k'\over(2\pi)^2}{e^{i({\bf k}+{\bf k}')\cdot
{\bf z}}\over(-i\omega_n-\xi_{k'})(i\omega_n-\xi_k)}\left[\vec{\Pi}^2
+\hbox{sin}2\theta\left(\Pi_x^2-\Pi_y^2\right)\right]\left[c_k\Delta^*_s
({\bf r})+e_k\Delta^*_d({\bf r})\right]\bigg\}; \label{deltasb_avh} \\
\Delta^*_d&=&{TV\over 2}\sum_{\omega_n}\int{d^2k\over(2\pi)^2}\bigg\{
{d_k\Delta^*_d({\bf r})\over\omega_n^2+\xi_k^2}
-{1\over 4(\omega_n^2+\xi_k^2)^2}\left[d_k^2|\Delta_d({\bf r})|^2
\Delta^*_d({\bf r})
+2c_kd_k|\Delta_s({\bf r})|^2\Delta^*_d({\bf r})+c_kd_k{\Delta^*_s}^2
({\bf r})\Delta_d({\bf r})\right]\nonumber \\
&-&\int d^2z{z^2\over 2}\int{d^2k'\over(2\pi)^2}{e^{i({\bf k}+{\bf k}')\cdot
{\bf z}}\over(-i\omega_n-\xi_{k'})(i\omega_n-\xi_k)}\left[\vec{\Pi}^2
+\hbox{sin}2\theta\left(\Pi_x^2-\Pi_y^2\right)\right]\left[e_k\Delta^*_s
({\bf r})+d_k\Delta^*_d({\bf r})\right]\bigg\}\label{deltadb_avh}, 
\end{eqnarray}

\noindent such that

\begin{eqnarray}
c_k&=&1+\hbox{cos}k_1\hbox{cos}k_2+\hbox{cos}k_1+\hbox{cos}k_2;\\
d_k&=&1+\hbox{cos}k_1\hbox{cos}k_2-\hbox{cos}k_1-\hbox{cos}k_2;\\
e_k&=&\hbox{sin}k_1\hbox{sin}k_2.
\end{eqnarray}

\noindent From these, and making use of (\ref{trick}), we can immediately
obtain the appropriate free energy (\ref{gl}) where

\begin{equation}
\{\alpha_s\;,\;\alpha_d\}={1\over V}-{1\over 4}\int{d^2k\over(2\pi)^2}
{\{c_k\;,\;d_k\} \over\xi_k}\hbox{tanh}\left({\xi_k\over 2T}\right),
\label{alpha_avh}
\end{equation}

\begin{equation}
\{\beta_1\;,\;\beta_2\;,\;\beta_3\;,\;\beta_4\}={T\over 16}\sum_{\omega_n}
\int{d^2k\over(2\pi)^2}{\{c_k^2\;,\;d_k^2\;,\;4c_kd_k\;,\;c_kd_k\}
\over(\omega_n^2+\xi_k^2)^2},
\end{equation}

\begin{equation}
\{\gamma_s\;,\;\gamma_d\}=T\sum_{\omega_n}\int{d^2k\over(2\pi)^2}{\hbox{sin}^2
k_1\left(t_{11}+2t_{20}\hbox{cos}k_2\right)^2+\hbox{sin}^2k_2\left(t_{11}
+2t_{20}\hbox{cos}k_1\right)^2\over\left(\omega_n^2+\xi_k^2\right)^2}\{c_k\;,
\;d_k\},
\label{gamma_avh}
\end{equation}

\begin{equation}
\gamma_{\nu}=-2T\sum_{\omega_n}\int{d^2k\over(2\pi)^2}{e_k^2
\left(t_{11}+2t_{20}\hbox{cos}k_1\right)\left(t_{11}
+2t_{20}\hbox{cos}k_2\right)\over\left(\omega_n^2+\xi_k^2\right)^2}.
\end{equation}

\noindent Note that $\beta_3=4\beta_4$, as was found for the EH 
model (\ref{gl_eh}) and in Ref.\ \onlinecite{ren}.
This generic feature of the free energy (\ref{gl}) is a direct 
consequence of the symmetry between
the s-wave and d-wave order parameters and the underlying bond (or 
directionally-dependent) gap functions from which they are derived.
Due to the complicated angular dependence of the AvH dispersion 
(\ref{disp_avh}), the above expressions above are not analytically tractable,
however. At optimal hole doping ($\langle n_{\rm hole}\rangle=1-\langle
n_{\rm electron}\rangle\sim 0.2$), the d-wave transition 
temperature found using Eq.\ (\ref{deltad_avh}) is $\sim$ 100 K, and the 
coefficients of the GL free energy have been evaluated numerically: 

\begin{eqnarray}
F_s&=&F_n-11.3\left(1-{T\over T_d}\right)|\Delta_d|^2+10.8|\Delta_s|^2
+38.7|\Delta_s|^4+5720
|\Delta_d|^4+603\Big[|\Delta_s|^2|\Delta_d|^2+{1\over 4}\left({\Delta^*_s}^2
\Delta_d^2+{\Delta^*_d}^2\Delta_s^2\right)\Big]\nonumber \\
&+&2.69|\vec{\Pi}\Delta_s|^2+7.38|\vec{\Pi}\Delta_d|^2+1.50
\Big[(\Pi_y\Delta_s)^*(\Pi_y\Delta_d)-(\Pi_x\Delta_s)^*(\Pi_x\Delta_d)+c.c\Big]
+{h^2\over 8\pi},
\label{gl_avh}
\end{eqnarray}

\noindent where the coefficients $\beta$, $\gamma$ have been evaluated at 
$T_d$. 

The maximum value of the s-wave component nucleated in the vortex core, 
calculated numerically using (\ref{maxsub}), (\ref{alpha_eh}), (\ref{gamma_eh}),
(\ref{alpha_avh}), and (\ref{gamma_avh}),
is shown as a function of hole density in Fig.\ \ref{max_all}. 
Evidently, in the AvH model the induced s-wave component can be at 
most $5\%$ of the bulk d-wave value (corresponding 
to $\langle n_h\rangle\sim 25\%$ and $T\rightarrow 0$, where it is assumed that
$\alpha_d(T)\propto (1-T/T_d), \forall T$). This is in spite of
a large temperature-independent gradient coefficient ratio 
($\gamma_{\nu}/\gamma_d\sim 20\%$ near optimal doping) that is found to
increase monotonically with doping up to the largest densities studied.
By contrast, with a suitable choice of parameters (low $T,V_0$, large 
$\langle n_h\rangle,V_1$), the EH model can allow 
for $s_{\rm max}/d_{\rm bulk}\sim 20-30\%$. As the on-site repulsion is
increased, however, the induced s-wave component is suppressed. The reduction 
in the magnitude of the s-wave component in the `overdoped' AvH model may be
partly due to effective on-site repulsion, built into the Hamiltonian by
constraining holes to move within a single spin sublattice of an 
antiferromagnetic background.

The GL d-wave coherence length $\xi(T)$, calculated numerically with the aid of
(\ref{cohere}), is shown as a function of hole concentration for $T=0.9T_d$ in 
Fig.\ \ref{coh_all}. As expected, $\xi(T)$ becomes shorter with decreasing 
$\langle n_h\rangle$. For hole densities up to $30\%$ filling, the AvH model
leads to $\xi\sim a/\sqrt{1-T/T_d}$ where $a$ is a lattice spacing,
in agreement with experiments for the cuprates.\cite{coherence} The EH model
yields similar d-wave coherence lengths for $V_1=2t,3t$. 
These short $\xi_d$ strictly violate the initial GL condition that the gap 
functions be more slowly varying than the Fermi 
wavelength $k_F^{-1}\sim 1a-2a$, the characteristic length scale of the 
fermionic excitations. 
The close agreement, however, between the results of
integrating Eqs.\ (\ref{gl}) and (\ref{current})\cite{berlinsk,ren,franz2} 
and numerical investigations of short-$\xi_d$ superconductors within 
Bogoliubov-De Gennes theory\cite{pekka} indicates that the GL equations
derived thus far capture at least some of the essential physics.

\section{Extension of the GL equations}

The current and magnetic field distribution 
around an isolated vortex of a d-wave superconductor is found to be fourfold 
symmetric whenever an s-wave component is induced in the vicinity of the vortex 
core.\cite{franz2,ren} It has been recently shown, however, that a similar
anisotropy results from the inclusion of higher-order d-wave gradient terms in
the GL free energy (\ref{gl}) even in the absence of an s-wave 
component.\cite{ichioka2} The connection between these observations can be
made apparent by integrating out of the free energy the degree of freedom 
associated with the subdominant order parameter. Given a bulk d-wave 
superconductor, for example, Eq.\ (\ref{gl}) can be minimized with respect
to $\Delta_s^*({\bf r})$, yielding to lowest order in $(1-T/T_d)$

\begin{equation}
\Delta_s({\bf r})=-{\gamma_{\nu}\over\alpha_s}(\Pi_y^2-\Pi_x^2)
\Delta_d({\bf r}).
\label{news}
\end{equation}

\noindent Note that since $\Delta_d({\bf r})$ varies on the length 
scale $\xi_d$, we immediately
obtain $\Delta_s\sim\gamma_{\nu}/\alpha_s\xi_d^2\sim(1-T/T_d)^{3/2}$ in 
agreement with (\ref{subdom}). Upon substitution of (\ref{news}) 
into (\ref{gl}), the leading correction due to the induced s-wave component
will be of the 
form $-{\gamma_{\nu}^2\over\alpha_s}|(\Pi_y^2-\Pi_x^2)\Delta_d|^2$. The free
energy for an inhomogeneous d-wave superconductor then becomes:

\begin{eqnarray}
F_s&=&F_n+\alpha_d|\Delta_d({\bf r})|^2+\beta_2|\Delta_d({\bf r})|^4
+\gamma_d|\vec{\Pi}\Delta_d({\bf r})|^2\nonumber \\
&+&\eta_d|\Delta_d({\bf r})\vec{\Pi}\Delta_d({\bf r})|^2
+\gamma_{d+}|\vec{\Pi}^2\Delta_d
({\bf r})|^2+\left[\gamma_{d-}-{\gamma_{\nu}^2\over\alpha_s}\right]
|(\Pi_y^2-\Pi_x^2)\Delta_d({\bf r})|^2+{h^2\over 8\pi},
\label{glnew}
\end{eqnarray}

\noindent which clearly
indicates that the induced s-wave order parameter breaks rotational symmetry
in precisely the same way as a fourth-order d-wave gradient term.

In order to assess the relative importance of the induced subcritical component
in eliciting anisotropy, the coefficients of the higher-order 
terms included in (\ref{gl}), and $\gamma_{d(s)-}$ in particular, must be 
determined microscopically. The derivation of 
$\eta_s$ and $\eta_d$ appearing in (\ref{f1s}) and (\ref{f1d}) requires 
extending the second term on the right hand side of Eq.\ (\ref{gap_eq}) to 
include gradients of the gap functions. Applying similar techniques to those
described in previous sections, one obtains for the EH model:

\begin{eqnarray}
\{\eta_s\;,\;\eta_d\}^{\rm EH}&=&-4T\sum_{\omega_n}
\int{d^2k\over(2\pi)^2}{\xi_k^2\left(\nabla_{k_x}\xi_k\right)^2\over\left(
\omega_n^2+\xi_k^2\right)^4}\{16{a_k'}^4\;,\;b_k^4\}\nonumber \\
&\approx&-{31\zeta(5)N'v_F^2\over 32\pi^4T^4}\left\{
\left({\mu\over D}+{\epsilon\over 4}\right)^4\;,\;{3\over 8}\left({\mu+D\over D}
\right)^4\right\},
\end{eqnarray}

\noindent where the analytical solution is valid in the continuum limit, i.e.\
at low electron densities. For the AvH model,

\begin{equation}
\{\eta_s\;,\;\eta_d\}^{\rm AvH}=-2T\sum_{\omega_n}\int{d^2k\over(2\pi)^2}
{\xi_k^2\left[\left(\nabla_{k_1}\xi_k\right)^2+\left(\nabla_{k_2}\xi_k\right)^2
\right]\over\left(\omega_n^2+\xi_k^2\right)^4}\{c_k^2\;,\;d_k^2\}.
\end{equation}

\noindent The coefficients of the fourth-order gradient terms can be derived 
by evaluating in the thermodynamic limit terms in (\ref{gaps_eh}) and 
(\ref{gaps_avh}) hitherto ignored. After some manipulation, one obtains

\begin{eqnarray}
\{\gamma_{s\pm}\;,\;\gamma_{d\pm}\}^{\rm EH}&=&
-{T\over 48}\sum_{\omega_n}\int d^2z\int{d^2k\,d^2k'\over(2\pi)^4}
{e^{i({\bf k}+{\bf k}')\cdot{\bf z}}\over(-i\omega_n-\xi_{k'})
(i\omega_n-\xi_k)}x^2\left(x^2\pm 3y^2\right)\left\{4{a_k'}^2\;,\;b_k^2\right\},
\nonumber \\
\{\gamma_{s\pm}\;,\;\gamma_{d\pm}\}^{\rm AvH}&=&-{T\over 48}\sum_{\omega_n}
\int d^2z\int{d^2k\,d^2k'\over(2\pi)^4}
\ldots\left(r_1+r_2\right)^2\left[\left(r_1+r_2\right)^2\pm 3\left(r_1-r_2
\right)^2\right]\left\{c_k\;,\;d_k\right\},
\end{eqnarray}

\noindent where the ellipsis represents the $k$-dependent part of the 
integrand. The fourth-order gradient terms can be evaluated analytically in
the EH model for weak to intermediate coupling and low densities:

\begin{equation}
\left\{\gamma_{s+}\;,\;\gamma_{d+}\;,\;\gamma_{s-}\;,\;\gamma_{d-}\right\}
\approx -{31\zeta(5)N'\over 256}\left({v_F\over\pi T}\right)^4\left\{
{3\over 2}\left(
{\mu\over D}+{\epsilon\over 4}\right)^2\;,\;{5\over 8}\left({\mu+D\over D}
\right)^2\;,\;0\;,\;{2\over 8}\left({\mu+D\over D}\right)^2\right\}.
\end{equation}

The analytical results obtained for the EH model indicate that all the 
higher-order terms in the free energy (\ref{f1s}) and (\ref{f1d}) have a 
negative sign compared
with the second-order terms. The overall sign of these higher-order terms 
is of no consequence, however, provided the order parameters
are sufficiently small and slowly varying; while both conditions will be 
satisfied near $T_c$, we also find
$\gamma_{s(d)+}/\gamma_{s(d)}\sim(v_F/\pi T)^2\rightarrow 0$ 
as $\mu\rightarrow -D$. At low densities, the EH model predicts a 
vanishing $\gamma_{s-}$ coefficient, in accordance with the expectation that the
free energy of a continuum s-wave superconductor should have spherical
symmetry. Furthermore, $\gamma_{d+}/\gamma_{d-}=5/2$, as was found by 
Ichioka {\em et al.}\cite{ichioka2}

The analytical results for the EH model cannot adequately determine whether
anisotropy in the free energy is primarily due to the existence of the 
subdominant order parameter or fourth-order gradients, however.
For a bulk s-wave superconductor, the induced d-wave order parameter 
clearly breaks circular symmetry since $\gamma_{\nu}\neq 0$ for $\mu>-D$ while
$\gamma_{s-}=0$. Yet stronger coupling or lattice effects could lead to 
a non-vanishing $\gamma_{s-}$ which could compete with 
the $\gamma_{\nu}^2/\alpha_d$ coefficient in the free energy (\ref{glnew}). 
Assuming that $\alpha_s\approx 1/V_1$, we find for a bulk d-wave superconductor

\begin{equation}
{\gamma_{\nu}^2\over\alpha_s\gamma_{d-}}\approx -{196\zeta(3)^2V_1N(0)\over
31\zeta(5)}\left({\mu\over D}+{\epsilon\over 4}\right)^2,
\end{equation}

\noindent which is of order $-V_1/t$. This result indicates that the 
`subcritical' coefficient $\gamma_{\nu}^2/\alpha_s$
and the `asymmetric' coefficient $\gamma_{d-}$ give 
similar contributions to the anisotropy. It should be emphasized, however, that
d-wave superconductivity is not favoured at densities for which these
analytical results are strictly valid. 

\begin{figure}[tbp]
\centerline{\psfig{figure=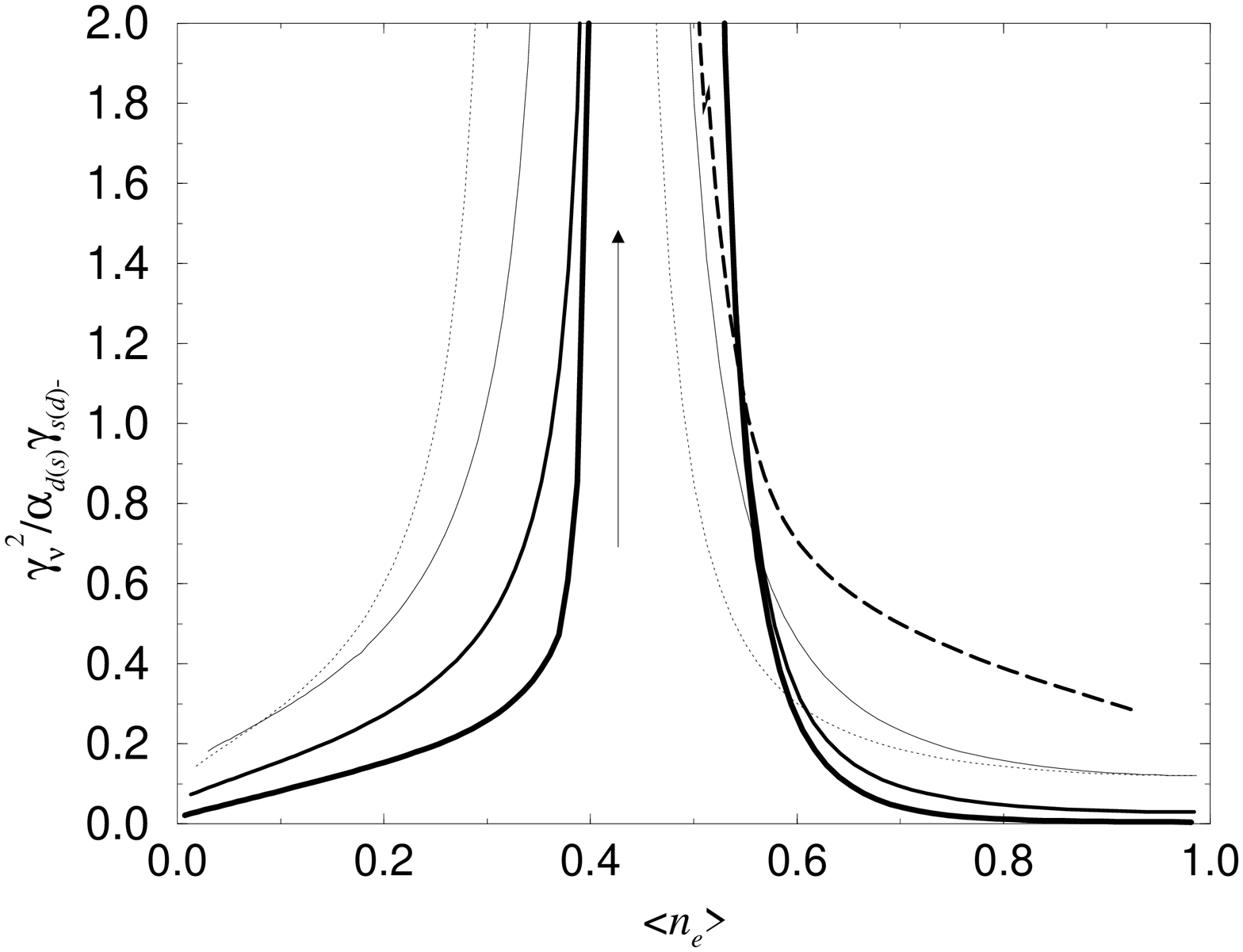,height=14.0cm,angle=270}}
\vskip1.0cm
\caption{The ratio $\gamma_{\nu}^2/\alpha_{d(s)}\gamma_{s(d)-}$ is shown as a
function of electron concentration $\langle n_e\rangle$ for $T=T_c$. Solid lines
(in order of decreasing boldness) correspond to the EH model for $V_1=1.3t$,
$2t$, $3t$ with $V_0=0$, while the dotted line is for the EH model with
$V_1=3t$, $V_0=4t$. The dashed line gives the results for the AvH model in the
electron notation such that $\langle n_e\rangle=1-\langle n_h\rangle$. To the
left (right) of the arrow is shown $\gamma_{\nu}^2/\alpha_d\gamma_{s-}$
($\gamma_{\nu}^2/\alpha_s\gamma_{d-}$) corresponding to bulk s-wave (d-wave)
superconductivity.}
\label{gad}
\end{figure}

The ratios $\gamma_{\nu}^2/\alpha_{d(s)}\gamma_{s(d)-}$ have been calculated
numerically for $T=T_c$ and are shown in Fig.\ \ref{gad}; virtually 
indistinguishable results have been obtained for all temperatures 
$0.5T_c\leq T\leq T_c$ (not shown). The numerics make
clear that the contribution to anisotropy of the subcritical coefficient is
at most comparable to the asymmetric coefficient for most 
densities and coupling strengths. Furthermore, since both coefficients have the
same sign, their contributions to the asymmetric gradient term in the free 
energy 
(\ref{glnew}) are in fact competing. Only for densities very near the crossover 
from bulk s-wave to d-wave condensation (or vice versa) in the EH model,
or for large hole densities in the AvH model, does the subcritical coefficient 
dominate; in this regime $\alpha\rightarrow 0$.

It is not presently clear if the overall sign of the anisotropic fourth-order
gradient term in the free energy (\ref{glnew}) has any physical significance. 
Previous
microscopic investigations of the EH model within the context of Bogoliubov-De 
Gennes theory\cite{pekka} demonstrated marked anisotropy in the structure of
the critical d-wave component near the core of an isolated vortex. Parameters
chosen correspond to a substantial s-wave component nucleated near the 
vortex core, i.e.\ $\gamma_{\nu}^2/\alpha_s\gg\gamma_{d-}$, and 
therefore a large negative coefficient for the asymmetric gradient term. Recent
work,\cite{markku} however, indicates that this anisotropy persists even for 
densities approaching half-filling, where $\gamma_{d-}$ dominates and the 
overall sign of the gradient term is positive. 

For completeness, the coefficients $\eta_{s(d)}$ and $\gamma_{s(d)+}$ have also
been determined numerically for both the EH and AvH models. Both coefficients
are always negative and in general it is found that for the EH model
$\eta_{s(d)}\sim 10\gamma_{s(d)+}\sim 100|\gamma_{s(d)-}|$, 
and $\gamma_{s(d)+}$ is of the same order as $\gamma_{s(d)}$ at $T_c$. For the 
AvH model at optimal doping and $T_c$, we obtain:

\begin{eqnarray}
F_s^{\rm AvH}&=&F_n-11.3\left(1-{T\over T_d}\right)|\Delta_d({\bf r})|^2
+5720|\Delta_d({\bf r})|^4+7.38|\vec{\Pi}\Delta_d({\bf r})|^2\nonumber \\
&-&16300|\Delta_d({\bf r})\vec{\Pi}\Delta_d({\bf r})|^2-16.5|\vec{\Pi}^2
\Delta_d({\bf r})|^2+2.65|(\Pi_y^2-\Pi_x^2)\Delta_d({\bf r})|^2
+{h^2\over 8\pi}.
\end{eqnarray}

\noindent The relatively large values of $\eta_{s(d)}$ and $\gamma_{s(d)+}$ 
compared with the magnitude of $\gamma_{s(d)}$ clearly demonstrates that the
GL theory derived herein is only strictly valid quite near $T_c$.

\section{Summary and Discussion}

The primary objective of the present work has been to derive 
the Ginzburg-Landau equations for a d-wave superconductor using two microscopic
lattice models which have been previously used to describe the high-$T_c$ 
oxides: the extended Hubbard model
and the Antiferromagnetic-van Hove model. In so doing, it has been
possible to quantitatively investigate how the lattice and external
magnetic field generate, and govern the interplay between, co-existing
s-wave and d-wave order parameters. In addition, the relative magnitudes of
the various GL coefficients, as well as their temperature and density 
dependence, have been ascertained.

Phenomenological GL theory has enabled much progress to be made recently
toward understanding the structures of isolated vortices and
the vortex lattice for d-wave superconductors. Of particular current interest
is the theoretical prediction by Franz {\em et al.}\cite{berlinsk,franz2}
of an oblique structure for the vortex 
lattice near $H_{c2}$, i.e.\ the Abrikosov lattice would be
intermediate between the usual triangle and a square.
The degree of `obliqueness' is mostly dependent on the gradient coefficient
ratio $\gamma_{\nu}/\gamma_d$. The coefficient $\gamma_{\nu}$ governs the 
extent an s-wave component is induced by spatial variations of the dominant 
d-wave order parameter, and characterizes the degree of fourfold symmetry in
the free energy. For $\gamma_{\nu}/\gamma_d=0$,
the s-wave component vanishes, yielding a triangular lattice. The Abrikosov
lattice deforms continuously away from a triangle as $\gamma_{\nu}/\gamma_d$ is
increased; for $\gamma_{\nu}/\gamma_d=0.45$, the angle between primitive 
vectors $\phi=76^{\circ}$. For $\gamma_{\nu}/\gamma_d\sim 0.6$ and higher, the 
flux lattice is square.

The present work demonstrates that microscopic models used to describe the
high-$T_c$ oxides can
predict a significant admixture of an s-wave order parameter in the mixed 
state of a d-wave superconductor. One consequence is the 
deviation of the flux lattice from that of a triangle.
It has been found that within a broad and experimentally relevant
parameter space both microscopic
models yield a gradient coefficient ratio $\gamma_{\nu}/\gamma_d\sim 0.1-0.4$.
This is consistent with two recent 
experimental observations\cite{keimer,maggio} for YBCO of flux lattices with
$\phi\approx 73^{\circ}$ and $77^{\circ}$. It is not yet clear, however, whether
the $a-b$ 
anisotropy associated with the orthorhombicity of YBCO is alone sufficient to 
account for the distortion in the flux lattice.\cite{timusk} 

It is presently uncertain whether the s-wave component nucleated in the 
vicinity of vortex cores is in fact required to induce a significant deviation
from a triangular vortex lattice. It has been shown that a
fourth-order gradient term in the d-wave order parameter also introduces
fourfold symmetry into the free energy; over much of the phase diagram, the
contribution of this term to anisotropy is comparable to that of the s-wave 
component. This important issue will addressed in future work.

Wherever possible, comparison has been made with previous derivations of
the GL coefficients within continuum models. It should be emphasized, however,
that lattice models not only provide considerably more information regarding 
the density and coupling-dependence of s-wave and d-wave admixture in the 
vortex core, but also avoid the technical difficulties (i.e.\ the application 
of the Pad\'e approximation) encountered in continuum models. In particular,
the lattice models clearly indicate that both s-wave and d-wave components must
always coexist in the vortex core for all temperatures below $T_c$, regardless
of the symmetry of the bulk order parameter.

\begin{acknowledgments}

The authors are grateful to W. Atkinson, A.J. Berlinsky, M. Franz, 
A. Nazarenko, C. O'Donovan, M.I. Salkola, and P.I. Soininen 
for numerous helpful comments and suggestions. This work has been
partially supported by the Natural Sciences and Engineering Research Council
of Canada and by the Ontario Centre for Materials Research.

\end{acknowledgments}

\appendix
\section*{Coefficients of the GL Gradient Terms}

The coefficients of the gradient terms appearing in the GL equations for the
gap functions (\ref{gaps_eh}) and (\ref{gaps_avh}) are, for the EH model:

\begin{eqnarray}
\epsilon_{x,1}^{\rm EH}&=&{1\over 2}|m|(|m|-1);\\
\epsilon_{y,1}^{\rm EH}&=&{1\over 2}|n|(|n|-1);\\
\epsilon_{x,2}^{\rm EH}&=&-{1\over 24}|m|(|m|-1)(|m|-2)(|m|-3);\\
\epsilon_{y,2}^{\rm EH}&=&-{1\over 24}|n|(|n|-1)(|n|-2)(|n|-3);\\
\epsilon_{xy,2}^{\rm EH}&=&-{1\over 4}|m|(|m|-1)|n|(|n|-1).
\end{eqnarray}

\noindent For the AvH model, we obtain:

\begin{eqnarray}
\epsilon_{x,1}^{\rm AvH}&=&{1\over 2}\Big[|m|(|m|-1)+|n|(|n|-1)+2mn\Big];\\
\epsilon_{y,1}^{\rm AvH}&=&{1\over 2}\Big[|m|(|m|-1)+|n|(|n|-1)-2mn\Big];\\
\epsilon_{x,2}^{\rm AvH}&=&-{1\over 24}\Big[|m|(|m|-1)(|m|-2)(|m|-3)
	+|n|(|n|-1)(|n|-2)(|n|-3)\Big]\nonumber \\
& &\quad -{1\over 12}|m||n|\Big\{3(|m|-1)(|n|-1)+2\Big[(|m|-1)(|m|-2)
+(|n|-1)(|n|-2)\Big]\Big\};\\
\epsilon_{y,2}^{\rm AvH}&=&-{1\over 24}\Big[|m|(|m|-1)(|m|-2)(|m|-3)
	+|n|(|n|-1)(|n|-2)(|n|-3)\Big]\nonumber \\
& &\quad -{1\over 12}|m||n|\Big\{3(|m|-1)(|n|-1)-2\Big[(|m|-1)(|m|-2)
+(|n|-1)(|n|-2)\Big]\Big\};\\
\epsilon_{xy,2}^{\rm AvH}&=&-{1\over 4}\Big[\Big(|m|(|m|-1)+|n|(|n|-1)\Big)^2
-4m^2n^2\Big].
\end{eqnarray}

\newpage

\begin{references}
\bibitem{invert}
The inversion of these equations requires the imposition of boundary 
conditions since the s-wave and d-wave gap functions on any given site are tied
to those at adjacent sites. Periodic boundary conditions are assumed throughout
the present work.
\end{references}
\end{document}